\documentclass[
floatfix,
aps,
prb,
amsmath,
twocolumn,
superscriptaddress
]{revtex4-1}
\usepackage{amsmath,amssymb}
\usepackage{amscd, latexsym}
\usepackage{mathrsfs}
\usepackage{graphicx}
\usepackage{epstopdf}
\usepackage{cancel}
\usepackage{amsfonts}
\usepackage{exscale}
\usepackage{dcolumn}
\usepackage{bm}
\usepackage{color}
\usepackage{natbib}
\usepackage{lmodern}
\usepackage[T1]{fontenc}
\usepackage[latin9]{inputenc}
\usepackage{verbatim}
\usepackage{float}
\usepackage {hyperref}
\hypersetup{colorlinks=true}

\usepackage[normalem]{ulem}

\usepackage[colorinlistoftodos]{todonotes}

\newcommand{\ket}[1]{\left| #1 \right>} 
\newcommand{\bra}[1]{\left\langle #1 \right|} 

\newcommand{\tr}[1]{{\rm Tr}\left\{ #1 \right\}} 

\DeclareMathOperator{\dif}{d\!}

\let\a=\alpha \let\b=\beta  \let\d=\partial 
  \let\q=\theta 

\newcommand{\diag}{{\rm diag}}

\def\be{\begin{equation}}
\def\ee{\end{equation}}
\def\bea{\begin{eqnarray}}
\def\eea{\end{eqnarray}}

\def\vep{{\varepsilon}}

\begin{document}

\title{Controlled--Z gate for transmon qubits coupled by semiconductor junctions}

\author{Zhenyi Qi}
\affiliation{Department of Physics, University of Wisconsin-Madison, Madison, WI 53706}
\author{Hong-Yi Xie}
\affiliation{Department of Physics, University of Wisconsin-Madison, Madison, WI 53706}
\author{Javad Shabani}
\affiliation{Department of Physics, New York University, New York, New York 10003}
\author{Vladimir E. Manucharyan }
\affiliation{Department of Physics, University of Maryland, College Park, MD 20742}
\author{Alex Levchenko}
\affiliation{Department of Physics, University of Wisconsin-Madison, Madison, WI 53706}
\author{Maxim G. Vavilov}
\affiliation{Department of Physics, University of Wisconsin-Madison, Madison, WI 53706}

\date{January 12, 2018}

\begin{abstract} 
We analyze the coupling of two qubits via an epitaxial semiconducting junction. In particular, we consider three configurations that include pairs of transmons or gatemons as well as gatemon-like two qubits formed by an epitaxial four-terminal junction. These three configurations provide an electrical control of the interaction between the qubits by applying voltage to a metallic gate near the semiconductor junction and can be utilized to naturally realize a controlled--Z gate (CZ). We calculate the fidelity and timing for such CZ gate. We demonstrate that in the absence of decoherence, the CZ gate can be performed under $50\ {\rm ns}$ with gate error below $10^{-4}$. 
\end{abstract}

\maketitle

\section{Introduction}
Over the last 10 years, superconducting circuits based on Al/AlO$_x$/Al tunnel Josephson Junctions (JJ) have clearly became a leading platform for implementing a solid-state based quantum computer~\cite{clarke2008superconducting, Lucero2008, dicarlo2009demonstration, Paik2011, devoret2013superconducting, wendin2017quantum}. Many factors contribute to this leadership. Absence of normal carriers helps to reduce decoherence; reproducibility of junction fabrication allows the pursuit of complex devices; the availability of quantum optics type of control of qubits, such as interactions of superconducting circuits with microwave cavities~\cite{yamamoto2003demonstration, chow2011simple, Plantenberg2007, Leek2009, Rigetti2010, Sheldon2016}, helps to achieve the highest degree of control at times even surpassing the benchmarks of conventional atomic systems. The most challenging task is to develop architectures that still maintain the high degree of control and a reasonable cost for scaling up. The scaling to a larger number of qubits meets technological challenges for current approaches with complicated circuitry necessary for individual control of large qubit systems.

A spectacular solution to scaling up system was found in conventional classical computers where successful simultaneous control of over a trillion transistors comes from the marvels of semiconductors. The transistor is the key element of this technology as it can switch on and off conductivity of a nanoscale region of a semiconductor chip by applying local nanosecond-scale voltage pulses. It is therefore tempting to explore an approach where highly-scalable gate-control of semiconductors can be incorporated into superconducting devices. A nanowire-based gatemon is a transmon whose tunnel Josephson junction is replaced with a hybrid super/semi--conductor junction. This junction can be formed by InAs nanowires in epitaxial contact with Al leads~\cite{chang2013tunneling,Lars2015,DeLange2015, Shabani2016, Doh272, Casparis2016, Kringhoj2017}.  The Josephson energy $E_J$ of such junctions is tuned by a side electric gate that controls the transmission of conducting channels in the semiconductor. The ability to tune the Josephson energy introduces Z gates for single qubit operations~\cite{Casparis2016}, permits additional reconfigurability of multiqubit systems to address frequency crowding and brings options for scaling quantum processor to a large number of qubits, by analogy with the operation of conventional transistor of classical processors. 

A universal quantum processor requires a combination of one and two qubit gates. One approach for two qubit gates was proposed for frequency tunable qubits by Strauch et al. \cite{Strauch2003}. This approach was demonstrated for transmon-like qubits~\cite{dicarlo2009demonstration,Bialczak2010,Yamamoto2010} and more recently for gatemons~\cite{Casparis2016}. However, as the number of qubits increases, their energy spectrum becomes very dense, making two-qubit gates based on frequency tuning a hardly scalable solution. 
A preferable approach would be to use tunable couplers between qubit pairs.

\begin{figure}
		%\centering
		\includegraphics[width=9cm]{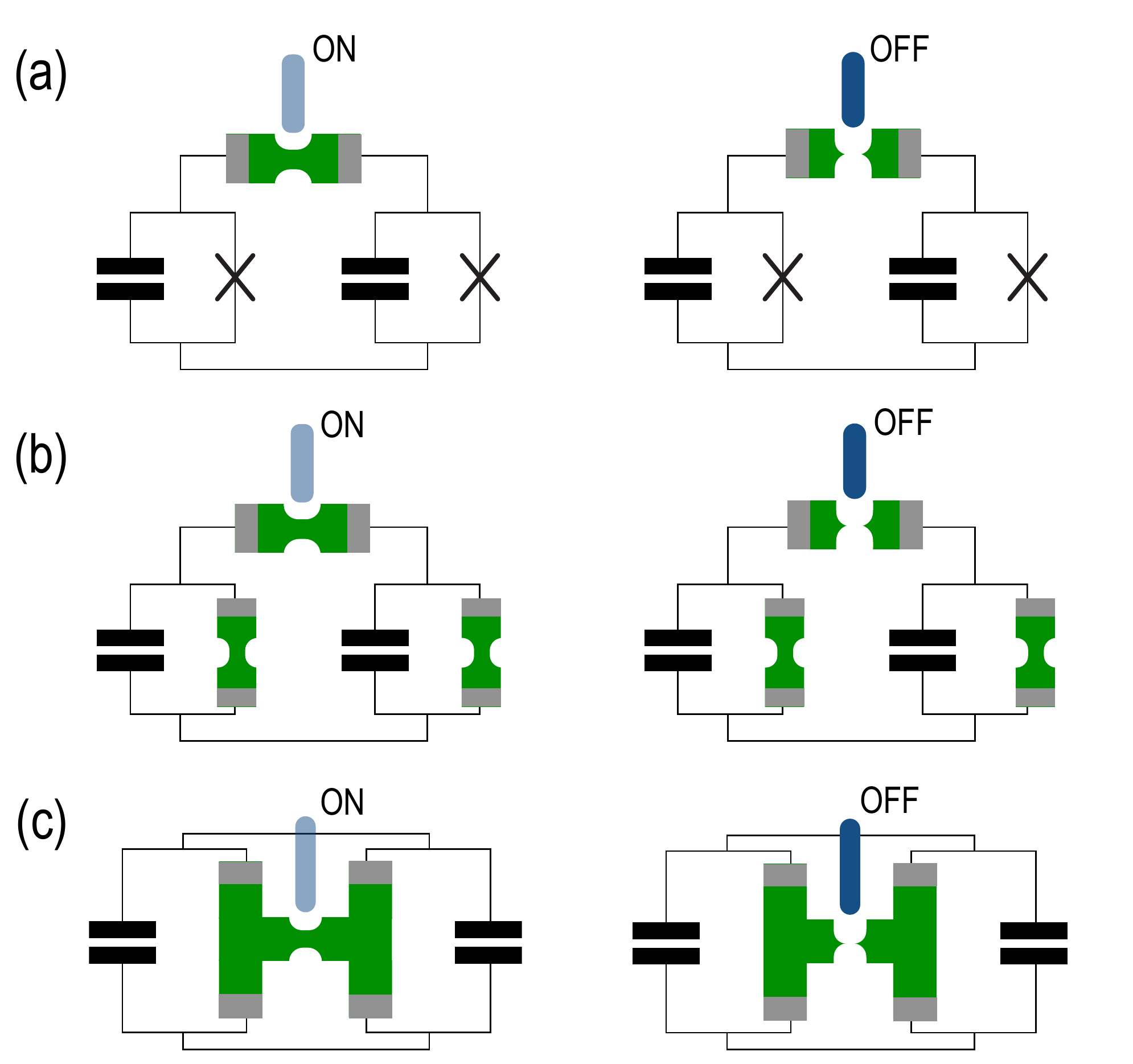}
		\includegraphics[width=7cm]{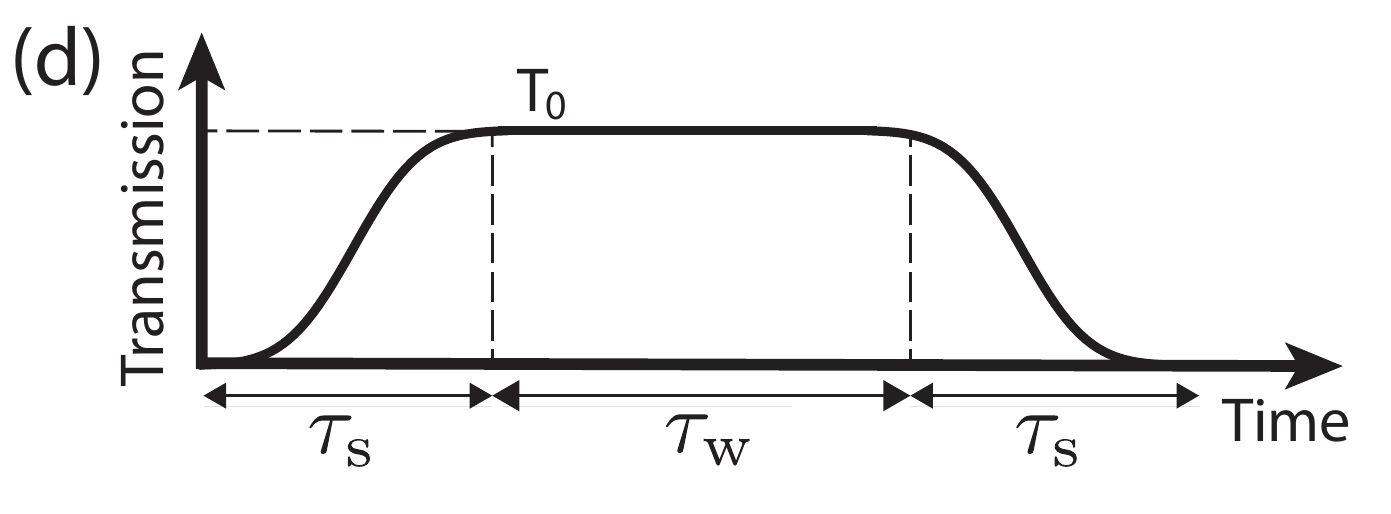}
		\caption{A schematic picture of two qubits connected by an epitaxial semiconductor junction (green bar). The three configurations presented here are: (a) two transmons formed by tunnel Josephson junctions, (b) two gatemons, where the Josephson elements are formed by other epitaxial semiconductor junctions, (c) two gatemon-like qubits, where their Josephson junction are combined into an H-shaped four-terminal epitaxial junction. The voltage on the gate (vertical bar) switches connection between two qubits on (left column) and off (right column). (d) The connector transmission $T_c$ as a function of time. The transmission changes according to Eqs.~\eqref{eq:Tonoff} during time $\tau_s$ for switching on and off  and is maintained at constant value $T_0$ during waiting time $\tau_w$.
		}
		\label{fig:transmon}
\end{figure}

In this paper, we propose a realization of a controlled--Z gate achieved by an electrically tuned semiconductor junction connecting two transmon qubits, see Fig.~\ref{fig:transmon} for an illustration. This approach is similar to the two qubit gates in g-mon systems~\cite{Chen2014a,Geller2015}, but in our case, the inductive coupling is controlled by the transparency of a semiconductor junction connecting two superconducting qubits. 

We assume that the connector can be completely closed by a proper negative voltage applied to a nearby gate. When the gate voltage is removed, an Andreev bound state (ABS) forms in the connector and its energy depends on the difference in phases of superconducting order parameters on the transmon superconducting islands. 
When the coupling between transmons through the junction is turned on, the additional Josephson energy associated with the ABS introduces effective interaction between two qubits and changes energy spectrum of the two qubit system. 
This change allows one to perform a controlled--Z gate (CZ), similar to frequency-tunable CZ gates~\cite{Strauch2003}.
The great advantage of this electro-statically controlled gate is that the interaction between the qubits can be turned off and thus completely decouple qubits. At the same time, the interaction can reach strong magnitude during the two-qubit gate operation necessary for a realization of fast high-fidelity CZ gates. Since the semiconductor junction is turned off during single qubit operations, we do  not expect that these junctions will degrade significantly coherent properties of the qubit system.

We have specifically discussed three qubit configurations:

\textit{Transmon} [Fig.~\ref{fig:transmon}(a)] is current favorite among superconducting qubits due to its reduction in sensitivity to charge noise relative to the Cooper pair box and increase in the qubit-photon coupling~\cite{Koch2007, majer2007coupling, dicarlo2009demonstration, Paik2011, rigetti2012superconducting}. We analyze the CZ gate characteristics for two transmons coupled by an electrically-controlled epitaxial semiconductor junction. This configuration will take advantage of coherence of conventional Al/AlO$_x$/Al transmons and electrically tunable interaction between them, see Sec.~\ref{Sec:transmon}.
	
\textit{Gatemon} [Fig.~\ref{fig:transmon}(b)] is a form of transmon, where the tunnel junction is replaced by a semiconductor Josephson junction~\cite{ Lars2015,DeLange2015, Shabani2016, Doh272, Casparis2016, Kringhoj2017}. The configuration of two gatemons coupled by a semiconductor junction has a benefit as this system has all-semiconductor system without AlO$_x$ tunnel junctions. However, always "on" semiconductor Josephson junction of a gatemon may reduce coherence of the individual qubits. We discussed this configuration in Sec.~\ref{sec:gatemon}.

\textit{H-pair} [Fig.~\ref{fig:transmon}(c)] is a system of two gatemon-like qubits with their Josephson junctions combined into a single H-shaped four-terminal junction which is lithographically patterned as epitaxial super/semi Josephson junctions~\cite{Shabani2016}.
The H-pair may simplify the fabrication process of coupled qubits. This junction also shows interesting physical structure of the Andreev bound states and was a focus of recent theoretical studies~\cite{van2014single,Nazarov-NC16,Meyer2017,Xie2017,Xie2017a, chang2013tunneling}. We describe properties of an H-pair in Sec.~\ref{sec:H-pair}. 

We demonstrate that the proposed CZ gate realization for all three configurations can reach fidelity above 99.99\% for $50\ {\rm ns}$ gate time.

% =======================================================
% =======================================================

%%========================================================================
%\begin{figure}
%	\centering
%
%	\caption{} 
%%formed by of the tunneling through connector during time $\tau_s$ and . We assume the time-dependence of wire transmission takes the form of .
%%$T_{c}^{(\rm on)}(t) = T_0 [{\rm erf}(4t/\tau_s-2)+1]/2$ in the switching-on stage and the switching-off is a time-reversed process $T_{c}^{(\rm off)}(t)=T_{c}^{(\rm on)}(\tau_s-t)$
%	\label{fig:pulse}
%\end{figure}
%%========================================================================

\section{Two coupled transmons\label{Sec:transmon}}

\subsection{Model of coupled transmons\label{subsec:TM_Model}}

We consider two transmon qubits coupled by a semiconductor junction. When the junction is open, the transmons are decoupled. Each transmon is characterized by the charging, $E_{C,\alpha}$, and Josephson, $E_{J,\alpha}$, energies, where $\alpha=1,2$. To suppress effects of charge noise on qubit, transmon capacitances $C_\alpha$ are chosen large and the charging energy $E_{C\alpha}=e^2/2C_\alpha$ are much smaller than the Josephson energy, $E_{C\alpha} \ll E_{J\alpha}$. 
The Hamiltonian of a transmon contains both charging and Josephson energies~\cite{Koch2007} and can be written in the form
\begin{eqnarray}
H_\alpha^{(0)} = 4 E_{C,\alpha} \hat n_\alpha^2+ E_{J,\alpha} \left(\frac{\hat \theta_\alpha^2}{2}-\frac{\hat \theta_\alpha^4}{24}
\right).
\label{eq:Htransmon}
\end{eqnarray}
Here, the electron number operator, $\hat n_\alpha$, and the superconducting order parameter phase operator, $\hat \theta_\alpha$, do not commute, $[\hat n_\alpha,\hat \theta_{\alpha}]=i$. 
The energy spectrum of the three lowest energy states of a transmon is characterized by the transition frequency $\omega_{10}$ between the ground and first excited states and the anharmonicity $\beta$:
\begin{eqnarray}
\hbar\omega^{(\alpha)}_{10} & = & E_{1}^{(\alpha)} - E_{0}^{(\alpha)}, \\
\hbar \beta^{(\alpha)} & = & E_{2}^{(\alpha)} - E_{1}^{(\alpha)} - \hbar\omega^{(\alpha)}_{10}.
\end{eqnarray}
These parameters are approximately given by the following expressions in terms of the Josephson and charging energies\cite{Koch2007}:
\begin{equation}
\hbar\omega_{10}^{(\alpha)} \approx \sqrt{8E_{J,\alpha} E_{C,\alpha}}-E_{C,\alpha},\quad
\beta^{(\alpha)} \approx E_{C, \alpha}/\hbar.
\label{eq:plasma_freq}
\end{equation}
While the condition $E_{J,\alpha}\gg E_{C, \alpha}$ is necessary to reduce transmon sensitivity to charge noise, this same condition results in weak anharmonicity of transmons and imposes certain constraints on time dependence of control pulses, including the CZ gate. 

In the discussion below, we assume that the Josephson energies of both transmons are equal, $E_{J,\alpha}=E_J$, but the charging energies $E_{C,\alpha}$ are different to provide distinguishable frequencies $\omega^{(\alpha)}_{10}$ by about 3\%, which requires $E_{C1}/E_{C2}\simeq 0.94$.
In particular, we take 
\begin{equation}
	\dfrac{E_J}{h}=20.55\ {\rm GHz}, \dfrac{E_{C, 1}}{h}= 240\ {\rm MHz}, \dfrac{E_{C, 2}}{h}= 255\ {\rm MHz}.\label{eq:Parameters}
\end{equation}
The qubit energies and anharmonicities for this choice of qubit parameters are presented in Table~\ref{tbl:energies}. 
%The energies of $\ket{11}$ state and $\ket{20} (\ket{02})$ state thus are 
%\begin{equation}
%	\dfrac{E_{11}}{h} = 12.22\ {\rm GHz},\dfrac{E_{02}}{h} = 11.75\ {\rm GHz}, \dfrac{E_{20}}{h} = 12.08\ {\rm GHz}\label{eq:E110220_TM}, 
%\end{equation}
%which 
The energy spectrum of two non-interacting qubits is such that computational state $\ket{11}$ is above both non-computational states $\ket{20}$ and $\ket{02}$. This order of energy states happens when the frequency separation between two qubits is smaller than the qubit anharmonicities. A counter example is considered in Sec.~III for the case of two gatemons, where the anharmonicity is reduced and the computational state $\ket{11}$ is between states $\ket{20}$ and $\ket{02}$.

%=======================================================================
% Transmon energy shift
\begin{figure}
	\centering
	\includegraphics[width=8cm]{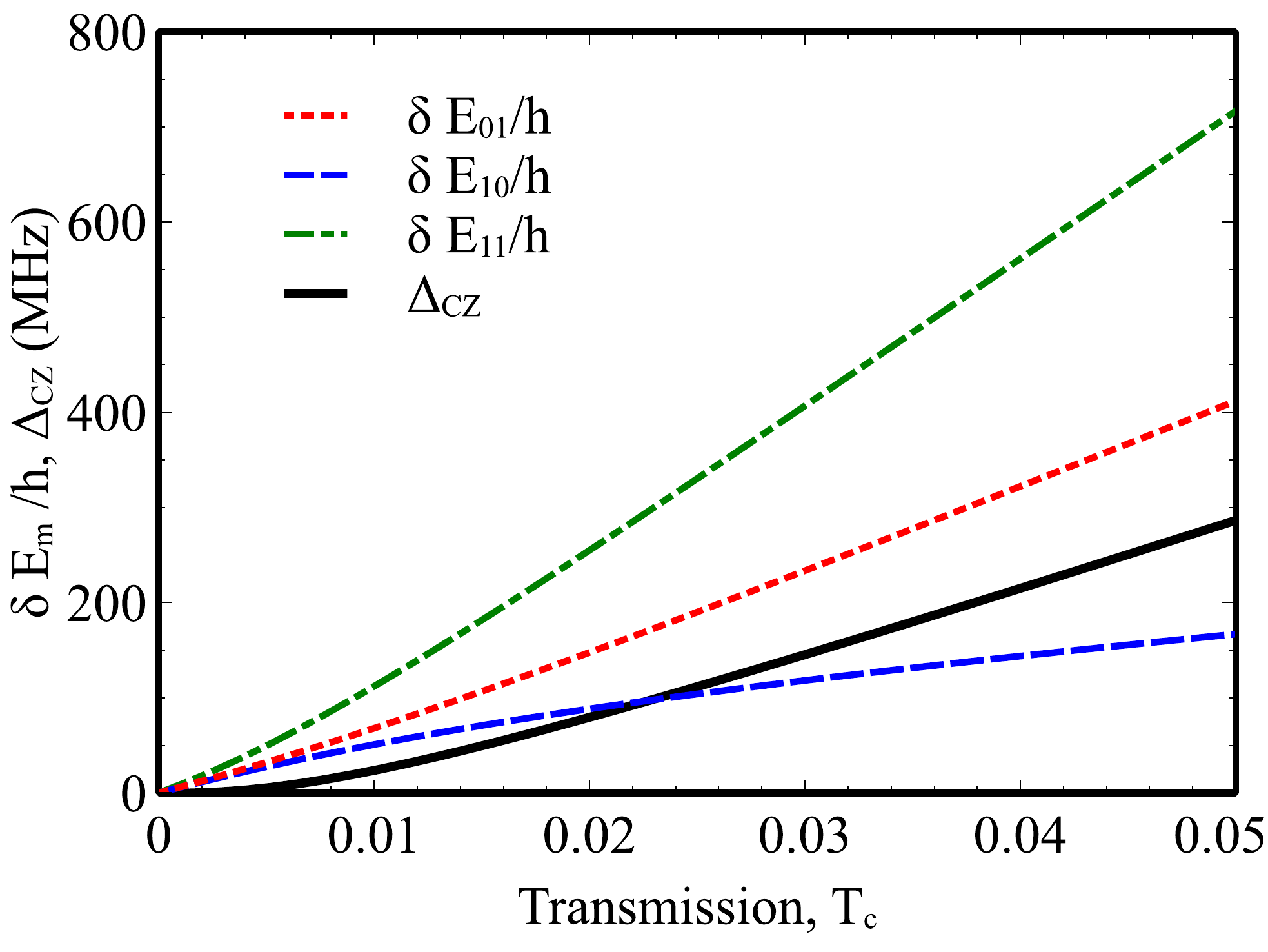}
	\caption{(Color online) Energy shifts on eigenenergies, $\delta E_{m}= \tilde E_m-E_m$, of the two-qubit systems as a function of transmission $T_c$. Only three out of four states in computational subspace are depicted. 
	%The shift $\delta E_{00}$, omitted for a clearer view, can be calculated using the relative shifts $\Delta_{\rm CZ} = (\tilde E_{11} +\tilde E_{00} -(\tilde E_{10}+\tilde E_{01}))/h$ shown as thick black solid line. 
	For both qubits, the Josephson energy $E_J/ h = 20.55\ {\rm GHz}$, and the charging energies $E_{C1}/ h = 240 \ {\rm MHz}$ and $E_{C2}/ h= 255\ {\rm MHz}$.} 
	\label{fig:spectrum_TM}
\end{figure}
%========================================================================

\begin{table}
\begin{tabular}{|c|c|c|c|}
\hline 
  & Transmon, GHz& Gatemon, GHz & H-junction, GHz \\ 
\hline 
$\omega_{10}^{(1)}/ h$ & 6.02 & 6.22 & 6.17\\ 
\hline 
$\omega_{10}^{(2)}/h$ & 6.20 & 6.41 & 6.36\\ 
\hline 
$\beta^{(1)}/ 2\pi$ & -0.294 & -0.063 & -0.066\\ 
\hline 
$\beta^{(2)}/ 2\pi$ & -0.315 & -0.067 & -0.070\\ 
\hline 
\end{tabular} 
\caption{Eigenenergies for non--interacting ($T_c=0$) two transmons, two gatemons and H-junction qubit systems. The Josephson energy $E_J/ h= 20.55\ {\rm GHz} $ for both transmon and gatemon qubits and $E_J/ h = 20.24\ {\rm GHz}$ for H-junction qubits, the charging energy for qubit one $E_{C1}/ h= 240\ {\rm MHz}$ and qubit two $E_{C2}/ h= 255\ {\rm MHz}$. All the eigenenergies are written in reference to the ground state energy.} \label{tbl:energies} 
\end{table}

When the connecting junction acquires non-zero transmission $T_{c}$, the interaction energy between the transmons is given by the energy of the Andreev bound state (ABS)\cite{Beenakker-PRL91}
\begin{eqnarray}
\delta E_{\rm ABS}=\Delta\sqrt{1-T_{c}\sin^2\left(\frac{\hat\theta_1-\hat\theta_2}{2}\right)},\label{eq:ABS}
\end{eqnarray}
where 
%$T_c$ is the transparency of the wire in the absence of gate voltage, 
$\hat\q_{1,2}$ are superconducting phases of the transmons and $\Delta$ is the superconducting energy gap. Assuming that the phase fluctuations of both transmons are small, 
$\q_{1,2}\ll 1$, we expand Eq.~\eqref{eq:ABS} to quartic terms in the qubit phases:
\begin{equation}
V_{\rm int}\simeq \frac{T_{c}\Delta }{8}\left(
(\hat\theta_1-\hat\theta_2)^2-\frac{1}{12}(\hat\theta_1-\hat\theta_2)^4
\right).
\label{eq:Vint}
\end{equation}
Here we consider small transmission coefficient $T_c\ll1$ of the connecting junction, sufficient for high-fidelity CZ gate.
This interaction modifies the instantaneous eigenstates $\widetilde{\ket{m}}$ and energies $\tilde{E}_{m}$ of the full system Hamiltonian
\begin{equation}
H=H_1+H_2+V_{\rm int}.
\label{eq:Hfull}
\end{equation}
As the transmission coefficient $T_{c}$ increases, the distance between neighboring energy states increases as a consequence of energy level repulsion. 

Below, we identify states $\widetilde{\ket{m}}$ by their adiabatic evolution as a function of $T_c$ from the non-interacting case, $T_c=0$, \textit{i.e.} index $m$ is composed of two integers $m\to (n_1,n_2)$, representing $n_\alpha$th excited state of non-interacting transmon $\alpha=1,2$.
%The dressed computational state $\widetilde{\ket{11}}$ remains above the dressed non-computational states $\widetilde{\ket{20}}$ and $\widetilde{\ket{02}}$. 
We evaluate the relative energy shifts $\delta \tilde E_m = \tilde E_m-E^{(1)}_{n_1}-E^{(2)}_{n_2}$ of computational subspace eigenstates, $m= \{00; 01; 10; 11\}$ as a function of transmission coefficient $T_c$, represented in Fig.~\ref{fig:spectrum_TM}.

The evolution operator in the computational subspace in the eigenstate basis $\widetilde{\ket{m}}$ has the form 
\begin{subequations}
\label{eq:CZ}
\begin{eqnarray}
W(t)  & = & {\rm diag} \{e^{-i\phi_{00}}; e^{-i\phi_{01}}; e^{-i\phi_{10}}; e^{-i\phi_{11}}\},\\
\phi_m(t) & = & \tilde E_m t/\hbar.
\end{eqnarray}
\end{subequations}
The interaction provides a non-zero value for the CZ gate rate\cite{Strauch2003,Chow2013,Ghosh2013}
\begin{equation}
\Delta_{\rm CZ} = (\tilde{E}_{11}+\tilde{E}_{00}-\tilde{E}_{01}-\tilde{E}_{10})/h,
\label{eq:DeltaCZ}
\end{equation}
so that after time $t= 1/(2\left|\Delta_{\rm CZ}\right|) $, the evolution operator is equivalent to the ideal CZ gate $U_{\rm{CZ}}={\rm diag}\{1;1;1;-1\}$ with the phase shifts $\phi_{10}-\phi_{00}$ and $\phi_{01}-\phi_{00}$ to be compensated by single qubit Z gates $U_{z1}$ and $U_{z2}$.
The dependence of $\Delta_{\rm CZ}$ on $T_c$ is shown in Fig.~\ref{fig:spectrum_TM} by a solid thick line. We demonstrate below that the energy shift $\Delta_{\rm CZ}$ is sufficient for performing gate operation over time of $(2\left|\Delta_{\rm CZ}\right|)^{-1}\lesssim 50$ns.

With tunable coupler between two qubits, CZ gate can be realized by simply switching interaction on over time $\tau_{s}$, waiting time $\tau_w \approx 1/(2\left|\Delta_{\rm CZ}\right|)$ and switching interaction off over time $\tau_{s}$. The total gate time is 
\begin{equation}
\tau_g=2\tau_{s}+\tau_w.
\end{equation} 
First, we analyze the phase accumulation and transition probabilities during switching processes. In particular, we demonstrate 
that the relatively small separation of the computational state $\widetilde{\ket{11}}$ from the leakage states $\widetilde{\ket{20}}$ and $\widetilde{\ket{02}}$ results in non-negligible leakage of the system state from the computational subspace due to transitions during switching on and off. 
Then, we describe the overall gate performance that combines the evolution of the system during switching on and off processes and waiting for $\tau_w$ at fixed $T_c$.

\subsection{Switching interaction on/off \label{sec:transition}}

We now consider the process of switching on and off interaction between the transmon qubits which is realized by electrically changing the transparency of the connector by tuning the voltage on the gate. 
Specifically, we assume that the connector transmission $T_c(t)$ during the switching on and off 
processes is
\begin{subequations}
\label{eq:Tonoff}
\begin{eqnarray}
T_{c}^{(\rm on)}(t) & = & T_{0} \frac{{\rm erf}(4t/\tau_{s}-2)+1}{2},\\
T_{c}^{(\rm off)}(t) & = & T_{c}^{(\rm on)}(\tau_s-t).
\end{eqnarray}
\end{subequations}
%where these switching processes are time-reversed with respect to each other.
During the switching on and off of the interaction between qubits, care is needed to avoid transitions from $\ket{11}$ state to double excited states $\ket{20}$ and $\ket{02}$, as well as transitions between states $\ket{01}$ and $\ket{10}$. We take $\tau_s$ to be longer than the inverse anharmonicities $1/\beta^{(1,2)}$ or qubits detuning, $1/|\omega_{10}^{(1)}-\omega_{10}^{(2)}|$ to suppress these transitions.

The evolution operator of the two-qubit system during the switching of interaction is a solution to the Schrodinger equation 
\begin{subequations}
\begin{eqnarray}
&& i\hbar \partial_t U(t) = H(t) U(t),
\label{eq:evolve}
\\ 
&&
H(t) = \sum_{\alpha=1,2} H_\alpha^{(0)}+V_{\rm int}(t),\label{eq:Hfull_t}
\end{eqnarray}
\end{subequations}
where the interaction $V_{\rm int}(t)$, see Eq.~\eqref{eq:Vint}, changes in time in response to the changing connector transmission $T_c(t)$. The evolution operators $U_{\rm on/off}(\tau_s)$ 
at the end of switching on and off processes are used below to compose the evolution operator for the whole gate and to evaluate the gate fidelity. 

In particular, the operators $U_{\rm on}(\tau_s)$ define the phase shifts $\varphi_m^{\rm (on)}$ of instantaneous eigenstates $\widetilde{\ket{m}}$ of the full Hamiltonian $H(t)$ that are utilized later to work out timing for the full CZ gate. We numerically evaluate $U_{\rm on/off}(\tau_s)$ by solving Eq.~\eqref{eq:evolve} 
in a $10\times 10$ Hilbert space in the basis of a harmonic oscillator wave functions and make sure that the low energy states are evaluated accurately for the actual Hamiltonian $H(t)$ of the system, Eq.~\eqref{eq:Hfull_t}. 
%We first numerically evaluate $U_{\rm on/off}$ for desirable switching time $\tau_{\rm s}$ that ensures acceptable leakage error and then 
Then, we analyze the $4\times 4$ matrix $[U_{\rm{on}}^{({\cal Q}_2)}]_{mm'} = \widetilde{\bra{m}} U_{\rm{on}}(\tau_s) \ket{m'}$ in the computational subspace ${\cal Q}_2$ of two qubits.
%which is essentially $U_{\rm{on}}(\tau_s)$ projected on the dressed computational subspace,  
%Here, $P$ is simply the projection operator from the physical qubit space to $2\times2$ computational Hilbert space at $T_c=0$, and 
%\mv{\textbf{It is a correct description for the numerical code, but is not correct in general.  Let us discuss.}}
%$\tilde{P} = \sum_{i}\ket{i}\bra{\tilde{i}}$ connects the interacting, $T_c = T_0$, and non-interacting eigenenergy state basis, with $i$ running through all vectors of the basis in the computational space. 
Matrix $[U_{\rm{on}}^{({\cal Q}_2)}]_{mn}$ defines the evolution of a state from the non-interacting computational subspace, $T_c=0$, to a final state, projected to the dressed computational subspace at $T_c=T_0$, where $m,n=\{00,01,10,11\}$. 
The diagonal elements of this matrix determine the phase factor accumulated by state $\widetilde{\ket{m}}$ during the switching process, $\varphi_{m}^{(\rm{on})} = -{\rm arg} \{[U_{\rm{on}}^{({\cal Q}_2)}]_{mm}\}$. 
We obtain the relative phase difference relevant for the CZ gate 
\begin{equation}
\label{eq:deltaPhiCZon}
\delta \Phi_{\rm CZ}^{(\rm{on})} = \varphi_{11}^{(\rm{on})}+\varphi_{00}^{(\rm{on})}-(\varphi_{10}^{(\rm{on})}+\varphi_{01}^{(\rm{on})}),
\end{equation}
which we utilize $\delta \Phi_{\rm CZ}^{(\rm{on})} $ in the next subsection to evaluate the full gate time. 

\begin{figure}
		\includegraphics[width=8cm]{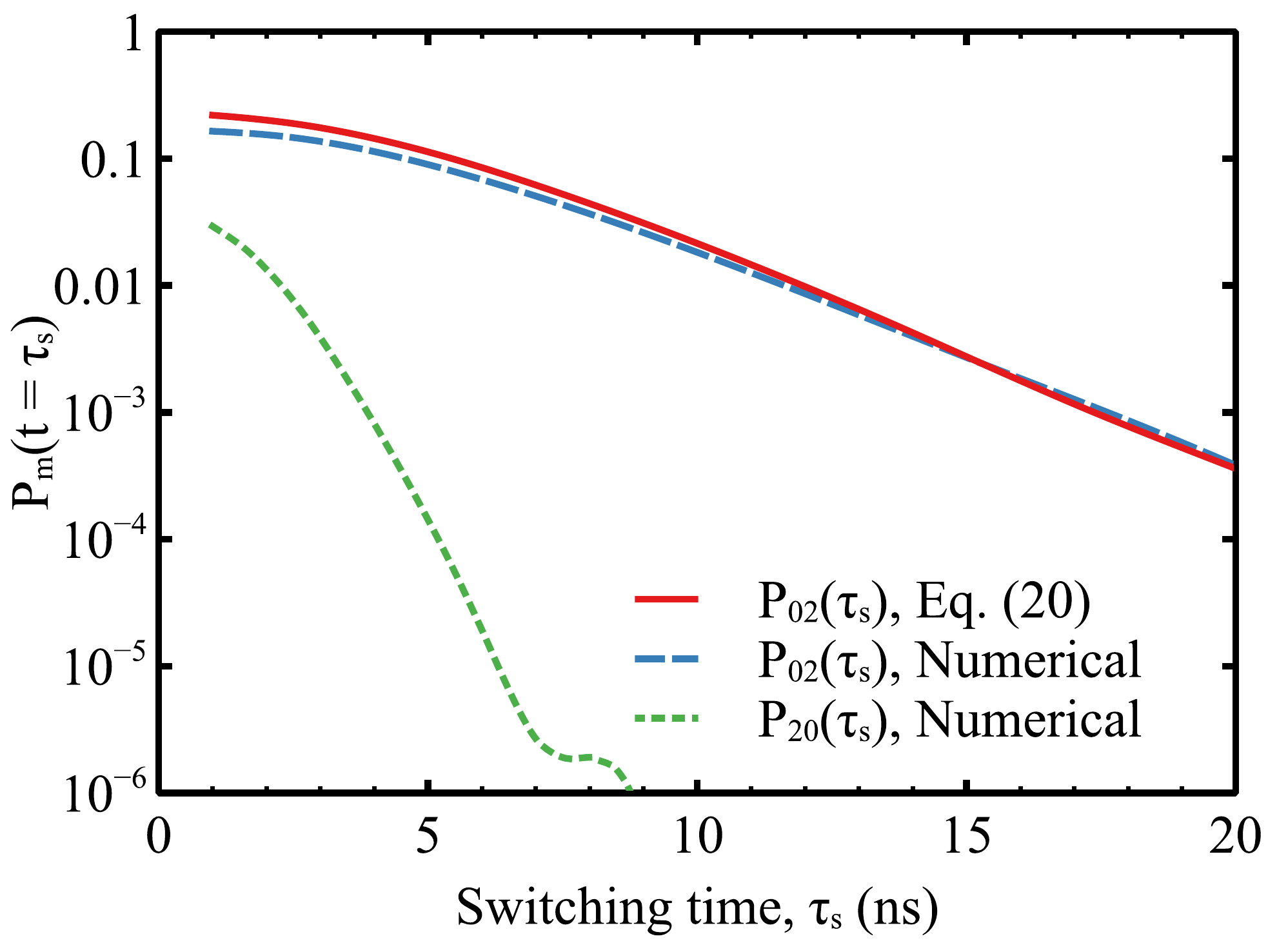}
		\caption{Leakage probability from state $\ket{11}$ to $\ket{02}$ and $\ket{20}$ during the switching on of transmission $T_c$ of a junction connecting two transmons as a function of switching time $\tau_s$ for the final transmission $T_0=0.015$. This leakage implies the worst case of gate performance since the transition from $\ket{11}$ to $\ket{02}$ is the dominant undesired transition. 
		}	
		\label{fig:Ptransitions}
\end{figure}

In the rest of this subsection, we evaluate the probabilities of transitions between pairs of instantaneous eigenstates since these transitions reduce the gate fidelity. 
We compute $[U_{\rm{on}}^{({\cal Q}_2)}]_{mn}$ and determine the probability of transition $P_{m,n}=|\widetilde{\bra{m}}U(\tau_s)\ket{n}|^2$ for a system of two transmons from state $\ket{n}$ to state $\widetilde{\ket{m}}$. The result of this calculation is shown in Fig.~\ref{fig:Ptransitions} by dashed and dotted lines for the transitions from state $\widetilde{\ket{11}}$ to its neighboring states $\widetilde{\ket{02}}$ and $\widetilde{\ket{20}}$. 
The probability of transitions 
between pairs of energy states decreases fast as the energy difference increases. 
For the energy spectrum of the two transmons analyzed here, the dominant leakage happens from  $\widetilde{\ket{11}}$ to $\widetilde{\ket{02}}$.
The transition probabilities also decrease fast as the switching time $\tau_s$ increases and drop below $10^{-3}$ for $\tau_s\gtrsim 20$ns.

%=======================================================================
%========================================================================

To better understand the transition probabilities between the states during the switching on and off processes, we perform the perturbation theory analysis following Ref.~\onlinecite{Ghosh2013}. In our case the bare level spacing is fixed and the interaction strength changes in time, while in Ref.~\onlinecite{Ghosh2013} the interaction was fixed and individual qubit spectrum was changing.
Denoting a dressed state of interest by $\widetilde{\ket{m(t)}}$, 
we can write $\widetilde{\ket{m(t)}} = \sum_n C_n(t) \exp(i\vartheta_n(t))\widetilde{\ket{n(t)}}$, where $\vartheta_n(t) = - \int_0^t \tilde{E}_n(t')\dif t'/\hbar$ and $\tilde{E}_n(t)$ is an eigenenergy of instantaneous eigenstate $\widetilde{\ket{n(t)}}$ at time $t$. The coefficients $C_n(t)$ are obtained as solutions to the Schr\"{o}dinger equation
\begin{equation}
\begin{split}
	\dot{C}_m(t) = & -C_m(0) \widetilde{\bra{m(t)}}\left[ \dfrac{\d }{\d t}\widetilde{\ket{m(t)}}\right] \\
	&- 
	\sum_n C_n(t) e^{i(\vartheta_n -\vartheta_m)} \dfrac{\widetilde{\bra{m(t) }}\dot{H}(t)\widetilde{\ket{n(t)}}}{\tilde{E}_n(t) - \tilde{E}_m(t)},
	\end{split}
	\label{Eq: dCdt_general}
\end{equation}
where $H(t)$ is the time-dependent Hamiltonian \eqref{eq:Hfull_t} of the system. 

Here we only provide perturbative analysis for the leakage from $\ket{11}$ and $\widetilde{\ket{02}}$, 
as these two states have the closest energy separation. Other pairs of neighboring states can be evaluated using the expressions obtained below with corresponding values for the instantaneous energy states and matrix elements of the interaction. The resulting transition probability from 
the initial state can be taken as the sum of probabilities of transitions to nearby states, provided that all these probabilities are small. 
We reduce our problem to the analysis of a two-level system formed by states $\ket{11}$ and $\ket{02}$: 
\begin{equation}
	H_{2L}(t) = \begin{pmatrix}
	E_{11}(t) & J(t)\\
	J(t) & E_{02}(t)
	\end{pmatrix}.
	\label{eq:H_2level}
\end{equation}
The time derivative of $H_{2L}$ is given by that of transmission $T_c$, which in general results in the time-dependence of the energies $E_{11/02}(t)$ and the interaction $J(t)$.

Starting from initial state $\ket{11}$, we take $C_{11}(t=0)=1$ and assume that the $C_{11}(t)\approx 1$ throughout the switching process.
The equation for $C_{02}(t)$ takes the form
\begin{equation}
 	\dot{C}_{02}(t) = -e^{i\chi(t)}
 	{\cal M}(t).
 	\label{eq:dCdt_1102}
\end{equation}
Here 
\begin{equation}
\chi(t) = \int_{0}^{t}
\sqrt{(E_{11}(t') - E_{02}(t'))^2 + 4J^2(t')}
\dif t'
\end{equation}
and the matrix element of the time-derivative of the Hamiltonian can be cast in the form
\begin{equation}
\begin{aligned}
&{\cal M}(t) = \dfrac{\widetilde{\bra{02(t) }}\dot{H}(t)\widetilde{\ket{11(t)}}}{\tilde{E}_{11}(t) - \tilde{E}_{02}(t)}\\
&=\dfrac{\dot{J}(t)(E_{11}(t) - E_{02}(t)) - (\dot{E}_{11}(t) - \dot{E}_{02}(t)) J(t)}{(E_{11}(t)-E_{02}(t))^2 + 4J^2(t)}.
\end{aligned}
\end{equation}
The transition probability $P_{02}(t)$ after time $t$ is determined by the integration
of Eq.~\eqref{eq:dCdt_1102} over time with the initial condition $C_{02}(t=0)=0$:
\begin{equation}
\begin{aligned}
C_{02}(t) =\int_{0}^t \dot{C}_{02}(t') dt',
\quad
P_{02}(t) = |C_{02}(t)|^2 .
\label{eq:P02}
\end{aligned}
\end{equation}
The perturbative analysis of the transition probability from $\ket{11}$ to $\widetilde{\ket{02}}$ is then accomplished by a numerical integration of Eq.~\eqref{eq:dCdt_1102}. The result is shown in Fig.~\ref{fig:Ptransitions}. Solid line in Fig.~\ref{fig:Ptransitions} represents $P_{02}(\tau_s)$ at the end of the switching on the connector transmission to $T_0=0.015$ during time $\tau_s$. Estimate~\eqref{eq:P02} captures the main features of the full numerical solution, to improve quantitative agreement between the curves, we would have to expand the above analysis to include multiple energy states into account. The structure of Eq.~\eqref{eq:P02} reveals that the suppression of $P_{02}(\tau_s)$ as a function of switching time is due to the fast oscillating factor $\exp(i\chi(t))$ in Eq.~\eqref{eq:dCdt_1102} for $\dot C_{02}(t)$, while ${\cal M}(t)$ is a smooth function of time.

\subsection{Controlled--Z gate\label{subsec:CZgate_TM}}

Now we analyze the dynamics of quantum states when the connector transmission is fixed, $T_c=T_0$, during time $\tau_w$. The phase difference combination for the CZ gate is given by $2\pi\Delta_{\rm CZ} \tau_w$ and the waiting time $\tau_w$ is found from the condition
\begin{equation}
\dfrac{2 \delta \Phi_{\rm CZ}^{\rm on}}{2\pi}+\Delta_{\rm CZ}\tau_w =\dfrac{1}{2},
\label{eq:phase_cond}
\end{equation}
where the phase difference $\delta \Phi_{\rm CZ}^{\rm on}$ was introduced in Sec.~\ref{sec:transition}.
To evaluate fidelity of the full gate, we numerically calculate the evolution operator $U$ for the process that is described by switching on transmission of the connector in the form of Eq.~\eqref{eq:Tonoff}, maintaining $T_c(t) = T_{0}$ for waiting time $\tau_w$ and switching off $T_{c}$ as a time-reversed process as illustrated in Fig.~\ref{fig:transmon}(d). 

For numerical evaluations, the corresponding evolution operator $U = U_{\rm{off}}U_{w}(\tau_w)U_{\rm{on}}$ is defined as the product of evolution operators $ U_{\rm{off}}$ and $U_{\rm{on}}$, discussed in Sec.~\ref{sec:transition} and the evolution operator $U_{w}(\tau_w)= \exp(-i H_{w} \tau_{w}/\hbar)$, where $H_{w}$ is the Hamiltonian in Eq.~\eqref{eq:Hfull} with $T_c = T_0$.
%\textit{cf.} Eq.~\eqref{eq:CZ},
Then, the fidelity is calculated by comparing matrix $[\hat W]_{nm}=\bra{n} U\ket{m}$ in computational $2\times2$ subspace, $n,m =\{00,01,10,11\}$, to the ideal CZ gate, $U_{\rm CZ}$, using the following expression for the fidelity~\cite{Pedersen2007}:
\begin{equation}
	F = \frac{1}{20}\left[ \tr{\hat W\hat W^\dagger} + 
	\left|\tr{ \hat U_{z1}\hat U_{z2}\hat  W\hat U_{\rm CZ}}\right|^2  \right],
\label{eq:fid}
\end{equation}
where $\hat U_{z1}$ and $\hat U_{z2}$ are single qubit gates such that 
$\hat U_{z1}\hat U_{z2} = {\rm diag}\{1, e^{i(\phi_{01} - \phi_{00})}, e^{i(\phi_{10}-\phi_{00})}, e^{i(\phi_{10}+\phi_{01}-2\phi_{00})}\}$, and $\phi_{m}$ are the phases of the diagonal elements of matrix  $\hat W$.

\begin{figure}
		\centering
		\includegraphics[width=8cm]{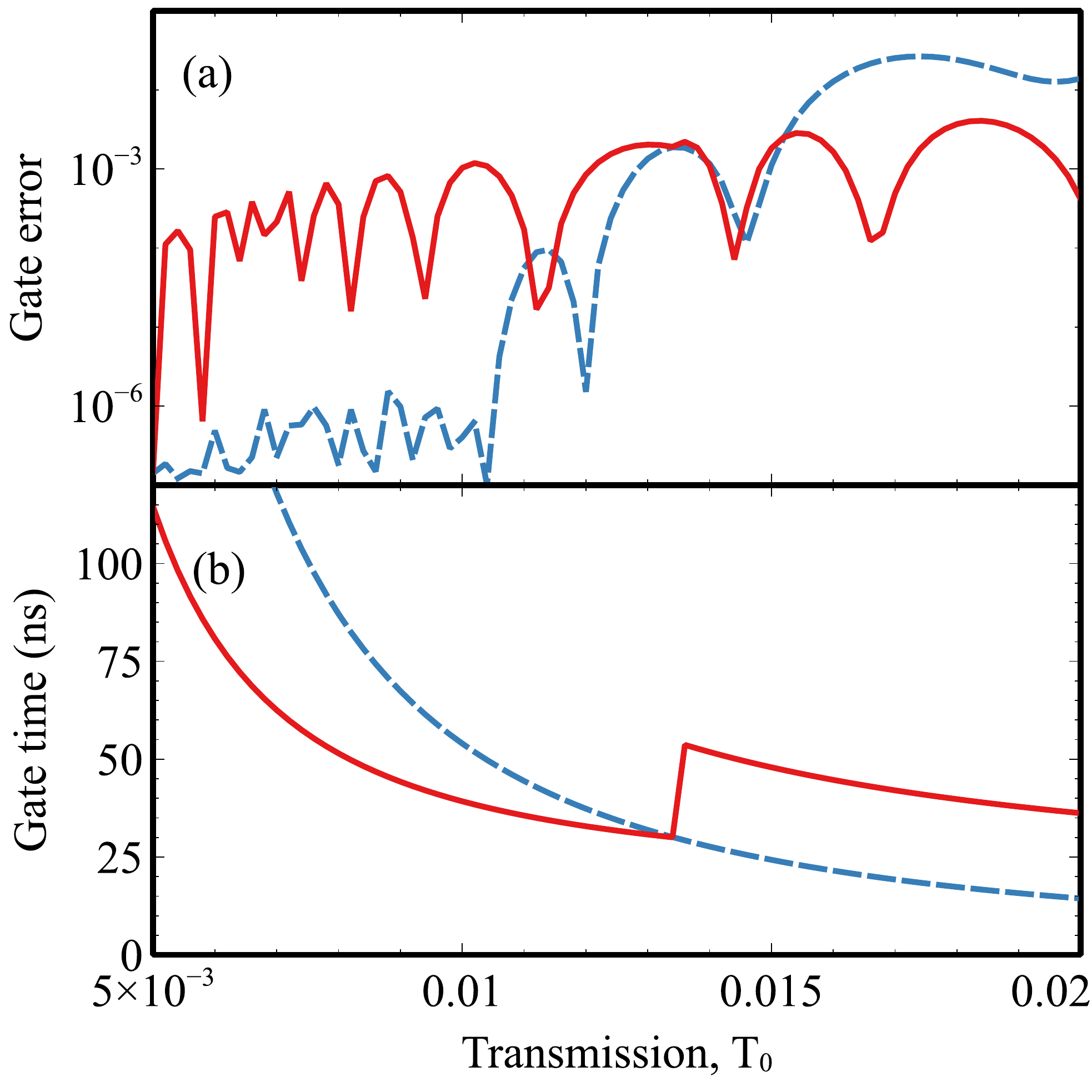}
		\caption{(a) (Color online) Error for the CZ gate and (b) corresponding gate time as functions of "on"-transmission $T_0$. The solid red lines are calculated assuming a pulse shape with an interacting plateau, so the gate time is $\tau_w+2\tau_s$. Here, we take $\tau_s=15\ {\rm ns}$. The dashed blue lines assumes the transmission is turned off right after it arrives its maximum value $T_0$, thus $\tau_w = 0$, and the gate time is $2\tau_s$, which is determined by the phase condition Eq.~\eqref{eq:phase_cond}. }
		\label{fig:Fig_cz}
\end{figure}

The gate errors of this voltage-controlled CZ gate, defined as $1-F$, are calculated as a function of  $T_0$, as shown in Fig.~\ref{fig:Fig_cz}(a). The error increases as $T_0$ becomes larger because for fixed switching time $\tau_{s} = 15\ {\rm ns}$, larger $T_0$ means the energy levels are shifted faster, which results in greater transition error, see Fig.~\ref{fig:Ptransitions}. The benefit of larger transmission $T_0$ is the shorter gate time because it requires less time for Eq.~\eqref{eq:phase_cond} to be achieved, as shown in Fig.~\ref{fig:Fig_cz}(b).  The trade-off between the gate fidelity and time determines the optimal value for $T_0$.
The kink of the red solid line, located at $T_{0} \simeq 0.013$, happens because the relative phase accumulated during the switching on and off processes equals $-\pi$, i.e., $\tau_{w} = 0$. Further increasing $T_{0}$ means the relative phase during the switching processes exceeds $-\pi$. To satisfy the phase condition \eqref{eq:phase_cond}, a finite-time interacting plateau is needed to complete another $2\pi$ rotation. The blue dashed lines represent the gate error and time of pulses whose interacting plateau is absent,  $\tau_{w} = 0$,  and the gate time $\tau_{\rm gate} = 2\tau_{\rm s}$. The time-averaged $T_c(t)$ is smaller for $\tau_w=0$ pulses, therefore longer gate times are required. At the same time, longer switching times allow for smaller gate errors. We observe that for small $T_0$ $(\le 0.01)$, the error can be reduced below $10^{-6}$ while the gate time is shorter than  $100\ {\rm ns}$.

%=======================================================================
% Transmon: Error and CZ gate time versus transmission
%========================================================================

\section{Two coupled gatemons\label{sec:gatemon}}

Replacing the insulating tunnel barrier between superconducting electrodes by semiconductor allows one to easily tune the Josephson Energy by using a electrostatic gate.~\cite{Lars2015} This type of transmon is thus named as gatemon for its gate tunable feature. With higher transparency between the tunnel barrier and the superconducting electrodes, expanding the ABS energy to the first order of $T_\alpha$ no longer well approximates the ABS energy, where $\alpha = 1, 2$ denotes the qubit index. Here, we consider single channel tunnel barriers with transmission $T_\alpha$ of gatemon $\alpha=1,2$. Then the Josephson energy of a qubit to the quartic order in $\theta_{\alpha}$ is 
\begin{equation}
	\begin{aligned}
	H_{J\alpha} & = -\Delta \sqrt{1 - T_{\alpha}\sin^2 \left(\dfrac{\hat \theta_{\alpha}}{2}\right)}\\
 				& \simeq -\Delta \left(1 - \dfrac{T_{\alpha}}{8}\hat\theta_{\alpha}^2 + \frac{T_{\alpha}}{96}\left(1-\dfrac{3T_{\alpha}}{4}\right)\hat\theta_{\alpha}^4\right).
	\end{aligned}
	\label{eq:gatemonHj}
\end{equation}
The anharmonicity of gatemon is suppressed~\cite{Kringhoj2017} by  $(1-3T_{\alpha}/4)$ in Eq.~\eqref{eq:gatemonHj}, which can usually be ignored in transmon systems with tunnel junctions containing many weakly transparent channels, $T_\alpha\ll 1$. 
We choose the parameters of the gatemon and transmon system to be identical and they are to be given by Eq.~\eqref{eq:Parameters}. The Josephson energy $E_J$ of a gatemons' single channel junctions corresponds to $T_{\alpha=1,2}=1$. The energy spectrum, including their anharmonicities are different for transmons and gatemons, see Table~\ref{tbl:energies}, due to the difference in coefficients of $\hat \theta_\alpha^4$ terms in Eqs.~\eqref{eq:Htransmon} and \eqref{eq:gatemonHj}.

%We also assume that $T_{\alpha, {\rm GM}} = \sum_{q} T_{q, \alpha, {\rm TM}} \simeq 1$, where the subscripts $\rm GM$ means the gatemon and $\rm TM$ means transmon, and the sum is taken over all channels for the transmon qubit. Anharmonicities and qubit frequencies are listed in the second column of 

\begin{figure}
	\centering
	\includegraphics[width=8cm]{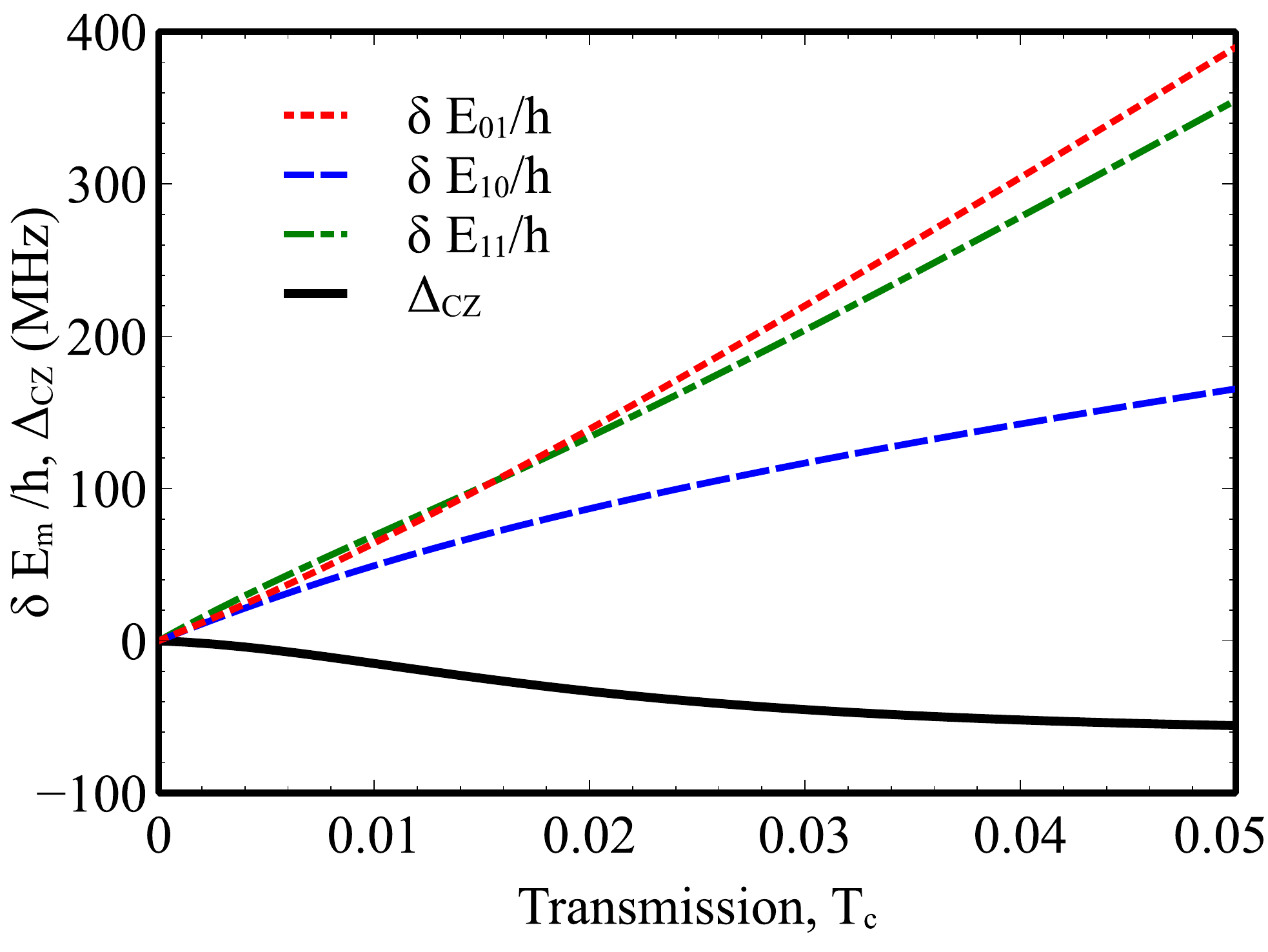}
	\caption{(Color online) Energy shifts on eigenenergies, $\delta E_{m}= \tilde E_m-E_m$, of the two-gatemon system as a function of transmission $T_c$. Only three out of four states in computational subspace are depicted. The shift $\delta E_{00}$, omitted for a clearer view, can be calculated using the relative shifts $\Delta_{\rm CZ} = (\tilde E_{11} +\tilde E_{00} -(\tilde E_{10}+\tilde E_{01}))/h$, shown as thick black solid line. The parameters are the same as the transmon system, as listed in Eq.~\eqref{eq:Parameters}.
	}
	\label{fig:spectrum_GM}
\end{figure}

The connecting junction between two qubits, once acquires non-zero transmission, $T_c\neq 0$, introduces an interaction between two qubits the same way as in transmon systems (see Sec.~\ref{subsec:TM_Model}). 
With reduced anharmonicity, the state $\widetilde{\ket{02}}$ has higher energy than its counterpart in transmon qubits. State $\widetilde{\ket{11}}$ is therefore sandwiched between $\widetilde{\ket{02}}$ and $\widetilde{\ket{20}}$ in this case. We also make a plot of the relative energy shift $\Delta_{\rm CZ}$ as a function of $T_c$ for the gatemon, shown in Fig.~\ref{fig:spectrum_GM}. As discussed in Sec.~\ref{subsec:CZgate_TM}, the non-zero $\Delta_{\rm CZ}$ enables the realization of CZ gate by switching on and off $T_c$, which inevitably introduces transition error. The most probable transition, the same as in transmon qubits, still happens between states $\widetilde{\ket{11}}$ and $\widetilde{\ket{02}}$ since this transition has smaller energy gap and larger interaction than other transitions involving computational basis. 
%=======================================================================
% Gatemon: energy shift
%========================================================================

To see how much the transition between states with smallest energy separation can reduce the gate fidelity during switching processes, we start from state $\widetilde{\ket{11}}$ and calculate the transition probability during switching-on process by (i) numerically calculating the evolution operator and (ii) approximating the system by a 2-level model and using Eq.~\eqref{eq:P02}. The 2-level approximation has been discussed in Sec.~\ref{sec:transition}. 

%========================================================================
% Gatemon: transition error.
\begin{figure}
	\includegraphics[width=8cm]{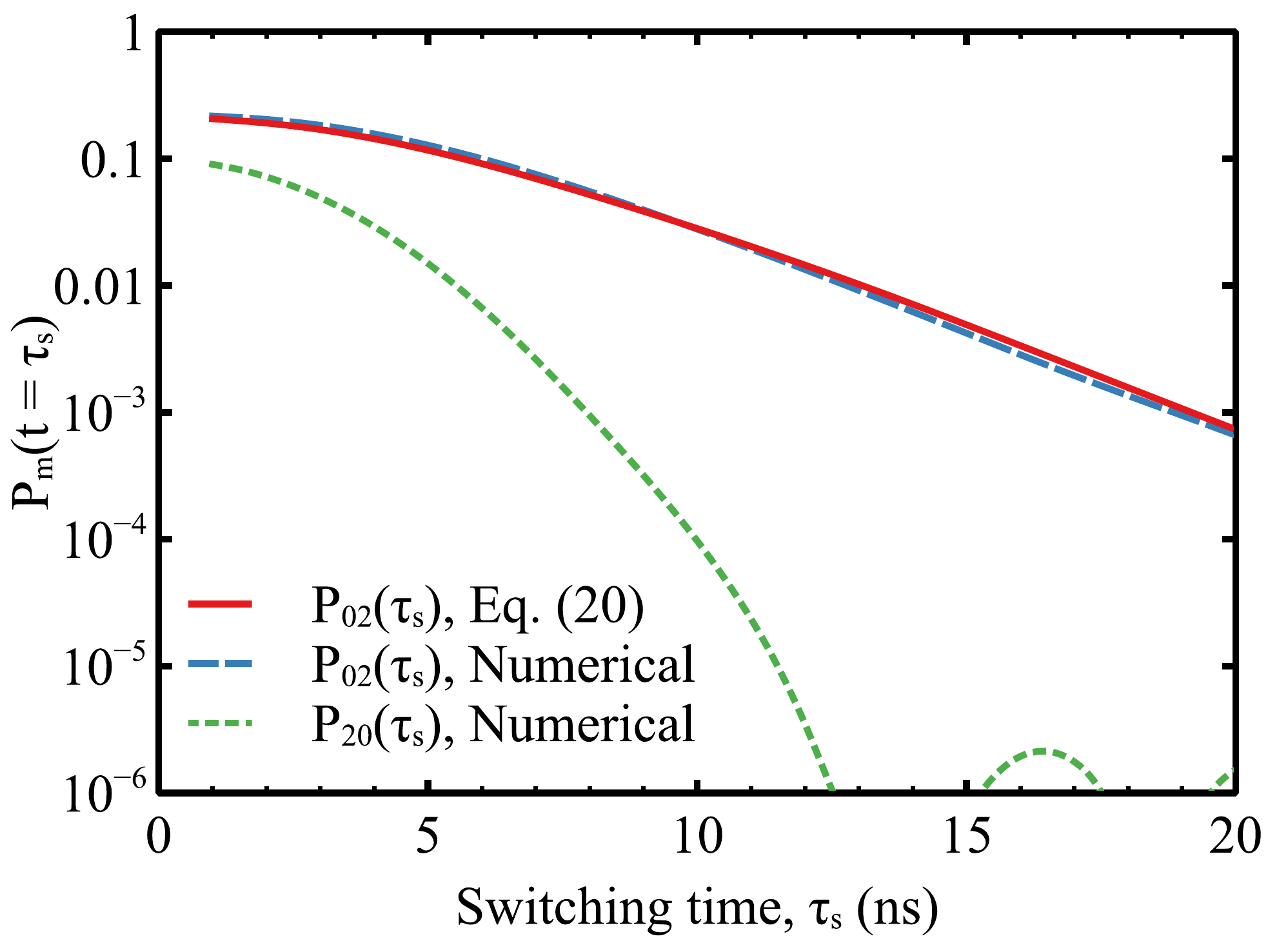}
	\caption{Leakage from state $\ket{11}$ to state $\ket{02}$ and $\ket{20}$ during the switching on of transmission $T_c$ of a junction connecting two gatemon qubits as a function of switching time $\tau_s$ for  $T_0=0.015$. 
	}
\label{fig:Ptransitions_GM}
\end{figure}

%========================================================================

As shown in Fig.~\ref{fig:Ptransitions_GM}, the transition probability from state $\ket{11}$ to state $\widetilde{\ket{02}}$ drops fast as the switching time increases. Specifically, the transition probability to state $\widetilde{\ket{02}}$ can be smaller than $10^{-3}$ when the switching time is larger than $20 \ {\rm ns}$.

%========================================================================
% Gatemon: gate time and gate error figures
\begin{figure}
	\centering
	\includegraphics[width=8cm]{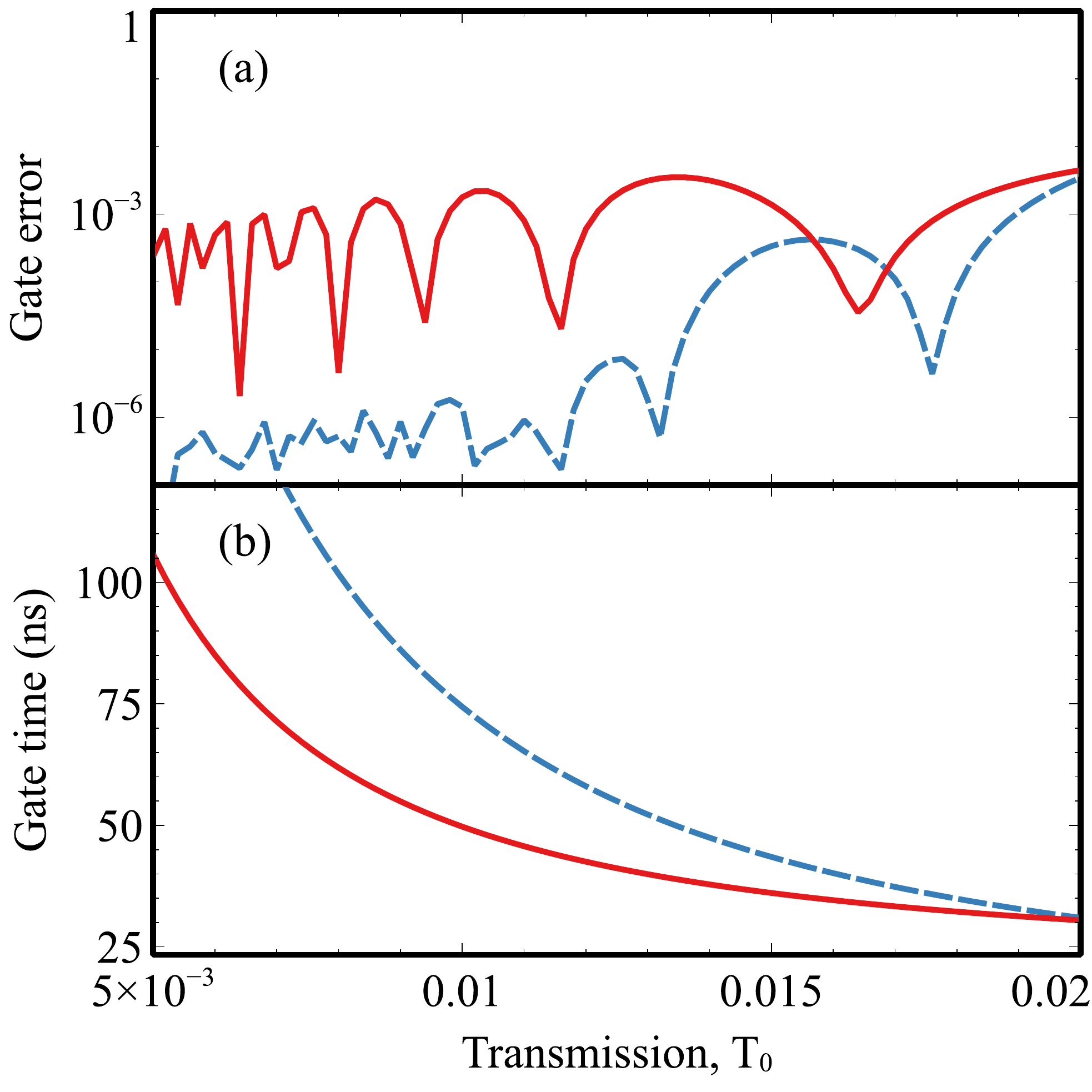}
	\caption{(a) (Color online) Error for the CZ gate on gatemon qubits and (b) corresponding gate time as functions of "on"-transmission $T_0$. The solid red lines are calculated assuming a pulse shape with an interacting plateau, so the gate time is $\tau_w+2\tau_s$. Here, $\tau_s = 15\ {\rm ns}$. The dashed blue lines assumes $\tau_w = 0$. }
	\label{fig:Fid_cz_GM}
\end{figure}
%========================================================================

 The relatively smaller shift of $\Delta_{\rm CZ}$ in Fig.~\ref{fig:spectrum_GM} indicates longer CZ gate time for gatemon than transmon qubits for the same $T_c$. And the decrease in anharmonicity results in potentially smaller fidelity for other qubit operations not discussed in this paper, since it makes the qubit transition frequencies among computational basis less distinguished from other undesired transitions involving levels out of the computational basis. Fig.~\ref{fig:Fid_cz_GM} shows both the gate error and gate time as functions of transmission $T_0$ for gatemon qubits. The solid lines are the gate error and gate time when finite-$\tau_w$ pulses are applied while the dashed lines correspond to $\tau_w=0$ pulses.
 
 Figure~\ref{fig:Fid_cz_GM}(a) does not show an obvious suppression in fidelity compared to the transmon qubits due to the reduced anharmonicity since the error in this CZ gate mainly comes from the transition error between states $\widetilde{\ket{11}}$ and $\widetilde{\ket{02}}$ during switching processes and no coherent drive is applied. The oscillations are a result of interference between states $\widetilde{\ket{11}}$ and $\widetilde{\ket{02}}$ during the gate. Fig.~\ref{fig:Fid_cz_GM}(b) shows that the gate time can be well below $50 \ {\rm ns}$ when $T_0>0.01$.

\section{Two qubits formed by a 4-terminal junction\label{sec:H-pair}}

\subsection{Josephson energy}
\begin{figure}[h!]
	 	 	\centering
	 	 	\includegraphics[width=6cm]{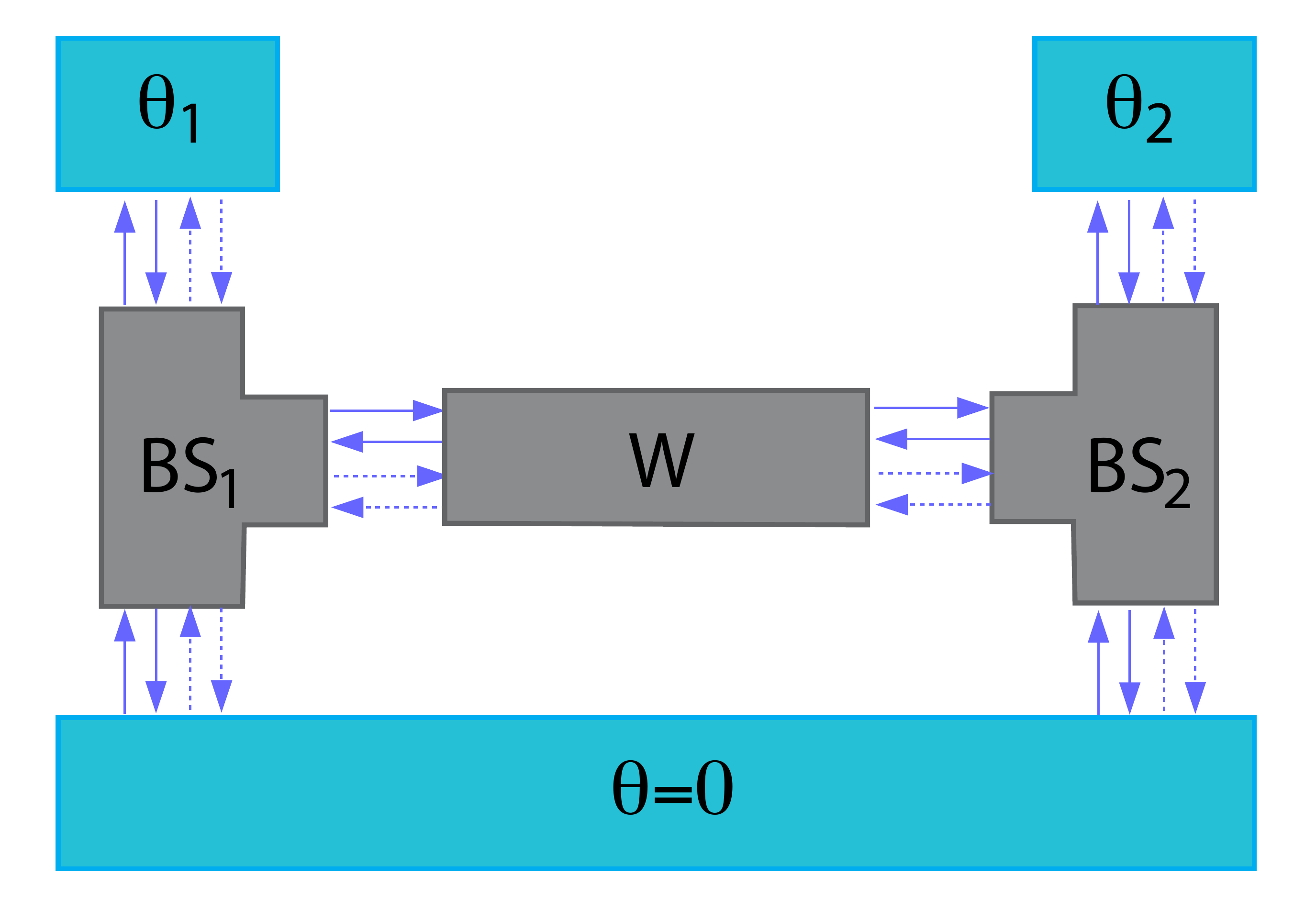}
	 	 	\caption{ H-junction consists of two semiconductor beam splitters, BS$_{1,2}$, connected by a short wire (W). The other two terminals of each beam splitter are attached to the ground superconducting strip and to the superconducting plates of the transmon--like qubit, with superconducting phase $\theta_{1,2}$.}
	 	 	\label{fig:Hjunction}
	 	 \end{figure}

We formulate a model for a system of two gatemon qubits connected through lithographically formed multiterminal junctions, such as an H--junction shown in Fig.~\ref{fig:Hjunction}. We assume that the junction is short, its scattering matrix $\hat S$ is energy independent and each terminal has only one conduction channel. Two terminals are connected to the superconducting ground lead, while the other two terminals are connected to the gatemon capacitor plates, see Fig.~\ref{fig:Hjunction}.

% =================================
% =================================

A set of localized sub-gap Andreev bound states are formed in the junction. These states are spin-degenerate which is guaranteed by Kramer's theorem. The energies of Andreev bound states are determined by the eigenvalue equation~ \cite{Beenakker-PRL91} 
\begin{equation}\label{Eq-Det}
\mathrm{det}\left[1-e^{-2i\chi}\hat{S}e^{i\hat{\theta}}\hat{S}^*e^{-i\hat{\theta}}\right]=0,
\end{equation} 
where the phase factor $\chi=\arccos(\epsilon/\Delta)$ captures Andreev electron-hole reflection at the superconductor-normal interface, $\Delta$ is the superconducting gap in the leads, and $e^{i\hat{\theta}}$ is a diagonal matrix that assigns the superconducting phase $\theta_i$ to each channel,
$\hat{\theta}=\diag \{\theta_1,\theta_2,\theta_3,\theta_4\}$. $\hat{S}$ is the scattering matrix of size $4\times 4$ for the H-junction, which is a combination of two three terminal junctions with two terminals connected by a short wire with transmission $T_c$; see the supplementary information. Since the phases of channels coupled to the ground superconducting lead can be fixed to zero, the Josephson energy is defined by Eq.~\eqref{Eq-Det} in terms of only two phase variables $\theta_{1,2}$. 

The Andreev energy states of the H-junction  are~\cite{Xie2017}
\be
\vep (\boldsymbol{\q}) = \pm \sqrt{\frac{A(\boldsymbol{\q}) + 4 \pm \sqrt{A^2(\boldsymbol{\q}) - 4 B(\boldsymbol{\q}) + 8}}{8}},
\label{eq:andreevE}
\ee
where $\boldsymbol{\q} \equiv (\q_1, \q_2 )$ and the $A$- and $B$-functions take the form
\begin{subequations} \label{A-B-qart}
\begin{align}
 A(\boldsymbol{\q}) = &\, A_0 + \sum_{\a=1}^{2} A_\a \cos{\q_\a} + A_{12} \cos(\q_1- \q_2), \label{A-qart} \\ 
 B(\boldsymbol{\q}) = &\, B_0 + \sum_{\a=1}^{2} B_\a \cos{\q_\a} + B_{12}^{-} \cos(\q_1 - \q_2)  \nonumber \\ 
   &\, + B_{12}^{+} \cos(\q_1 + \q_2). \label{B-qart}
\end{align}
\end{subequations}
In terms of the matrix elements of $\hat{S}$ the coefficients in Eqs.~(\ref{A-qart}) and (\ref{B-qart}) read
\begin{align}
& A_0 = 2 | S_{14} |^2 +\sum_{\b=1}^4 | S_{\b \b} |^2 , \quad A_{12} = 2 | S_{23} |^2, \nonumber \\
& A_\a = 2 \left( | S_{1,\a+1} |^2 + | S_{\a+1,4} |^2 \right), 
\end{align}
and 
\be
\begin{split}
\label{eq:Bs}
B_0 = &\, 2 \left( | S_{12} S_{24} -S_{14} S_{22} |^2 + | S_{13} S_{34} - S_{14} S_{33} |^2 \right) \\ 
      &\, +\sum_{\a<\b} \left| S_{\a\a} S_{\b\b} -S_{\a \b}^2 \right|^2, \\
B_1 = &\, | S_{13} S_{23} - S_{12} S_{33} |^2 + | S_{14} S_{24} - S_{12} S_{44} |^2 \\
      &\, + | S_{12} S_{14} - S_{11} S_{24} |^2 + | S_{24} S_{33} - S_{23} S_{34} |^2, \\
B_2 = &\, | S_{12} S_{23} - S_{13} S_{22} |^2 + | S_{14} S_{34} - S_{13} S_{44} |^2 \\
      &\, + | S_{13} S_{14} - S_{11} S_{34} |^2 + | S_{23} S_{24} - S_{22} S_{34} |^2, \\
B_{12}^{-} = &\, | S_{12} S_{13} - S_{11} S_{23} |^2 + | S_{24} S_{34} -S_{23} S_{44} |^2 \\
             &\, + | S_{14} S_{23} - S_{12} S_{34} |^2 + | S_{14} S_{23} - S_{13} S_{24} |^2, \\
B_{12}^{+} = &\, | S_{13} S_{24} - S_{12} S_{34}|^2.
\end{split}
\ee

The Josephson energy of an H-junction 
has a series expansion to the fourth order in the superconducting phases $\theta_{1,2}$:
\begin{equation}
H_{\rm JJ}(\theta_1,\theta_2)=\Delta
\sum_{i,j=0}^4 K_{ij}\theta_1^i\theta_2^j,
\label{eq:EJJ}
\end{equation}
where coefficients $K_{ij}$ are obtained from the expansion of Eq.~\eqref{eq:andreevE} to the forth order in $\theta_{1,2}$, and we take $K_{ij}=0$ for $i+j>4$ to avoid higher order terms. For $T_c=0$, the two qubits do not interact and 
\begin{equation}
H_{\rm JJ}(\theta_1,\theta_2)=\Delta \left(K^{(0)}_{20}\theta_1^2+K^{(0)}_{40}\theta_1^4+
K^{(0)}_{02}\theta_2^2+K^{(0)}_{04}\theta_2^4\right).
\label{eq:EJJnoint}
\end{equation} 
The full Hamiltonian of the system is similar to the Hamiltonian of a transmon Hamiltonian, Eq.~\eqref{eq:Htransmon} with $E_{J,1}=2 \Delta K^{(0)}_{20}$, $E_{J,2}=2 \Delta K^{(0)}_{02}$ and a modified anharmonicity. When the wire acquires finite transmission, $T_c\neq 0$, the interaction between the qubits develops with the interaction term given by
\begin{equation}
V_{\rm int} = \Delta
\sum_{i,j=0}^4 \delta K_{ij}\theta_1^i\theta_2^j,\quad \delta K_{ij} = K_{ij} -K^{(0)}_{ij}.
\label{eq:VJJ}
\end{equation}

The scattering matrix for the H-junction, Fig.~\ref{fig:Hjunction}, can be constructed in terms of the scattering matrices for each beam splitter and the connecting wire, see the  Appendix.
We use Eqs.~\eqref{eq:andreevE}-\eqref{eq:Bs} to characterize the qubit system, called H-pair, and evaluate the CZ gate fidelity for a particular choice of parameters for the beam splitters. We demonstrate that this system is sufficient for control of energy splitting of individual qubits and fast switching the interaction between qubits. 
%mv removed: We emphasize the present analysis is applicable to more general two-qubit configurations where the transmon-like qubits are connected by an H-junction. 

\subsection{Controlled--Z gate}

For the H-pair, state $\widetilde{|11\rangle}$ is close to leakage states $\widetilde{|20\rangle}$ and $\widetilde{|02\rangle}$. To reduce the excitations to non-computational states, we keep the interaction strength below the anharmonicity level, $\sim E_C$, and imply that the transmission of the connector is small, $T_c\lesssim \sqrt{E_c/\Delta}\ll 1$. 
Here we allow for the wire transmission to have larger values $T_c\simeq 1$, but we choose such parameters of the Y-junction that the resulting conductance of the H-junction between the qubits is small.
The Josephson energy, Eq.~\eqref{eq:EJJ}, acquires terms with small $K_{ij}\neq 0$ for both $i,j\neq 0$, resulting in interaction between the qubits.
In particular, the coefficient $K_{11}$ is the dominant term describing the coupling between the qubits.

%========================================================================
% gatemon gate time and gate error figures
\begin{figure}
	\centering
	\includegraphics[width=8cm]{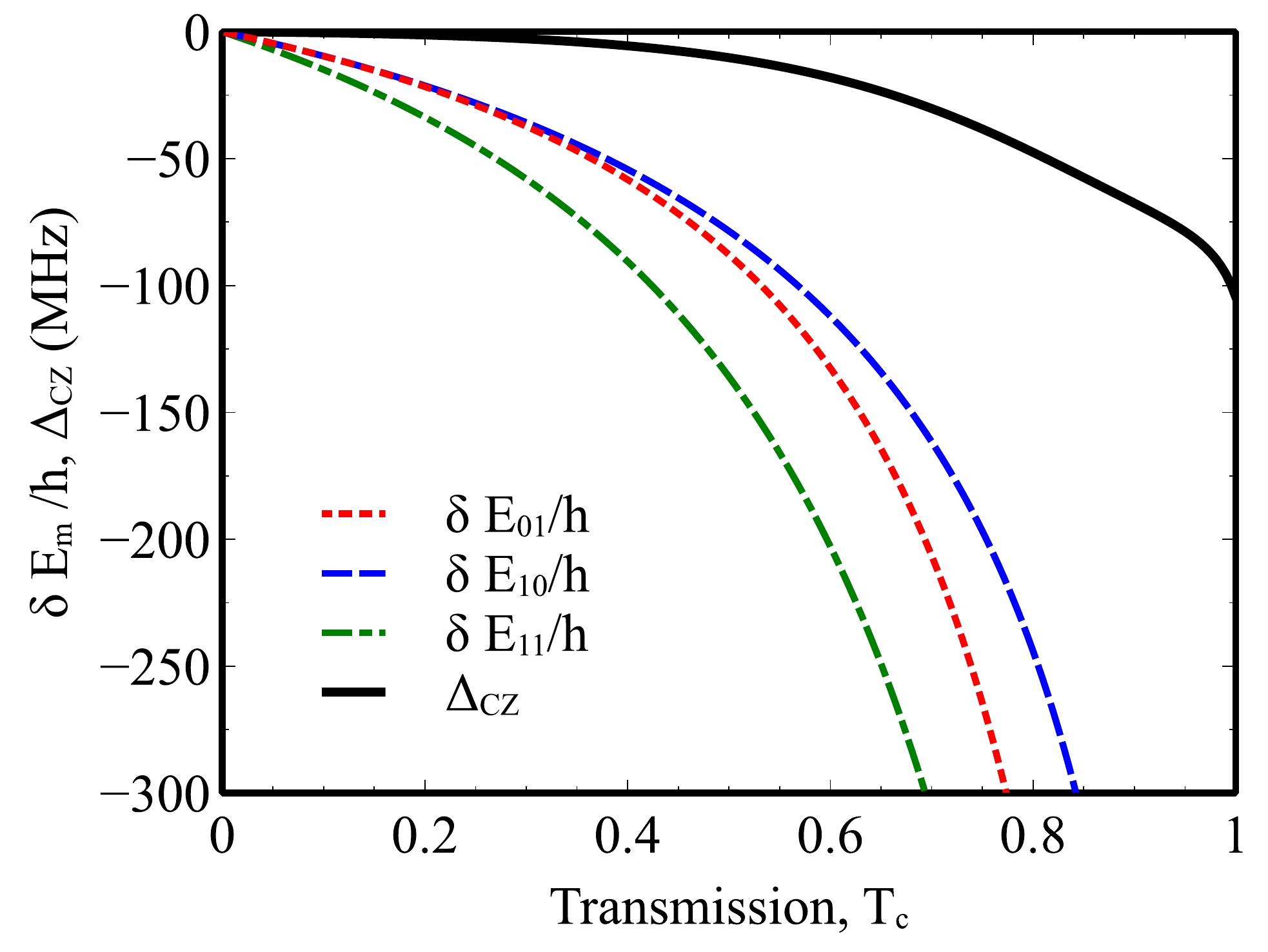}
	\caption{Energy shifts on eigenenergies, $\delta E_{m}= \tilde E_m-E_m$, of the two-qubit systems. The charging energies $E_{c1}/h = 240 \ {\rm MHz}$ and $E_{C2}/h= 255\ {\rm MHz}$ for qubit one and qubit two respectively, the same as the values given in Eq.~\eqref{eq:Parameters}. $K_{02, 20} = 0.123$, corresponding to Josephson energy $E_J/h = 20.24\ {\rm GHz}$.}
	\label{fig: energy_diagram_JQ}
\end{figure}
%========================================================================

The anharmonicity of the  H-pair is reduced by a numerical factor as compared to conventional AlO$_x$ oxide tunnel Josephson junction~\cite{Kringhoj2017}, as we also discussed in Sec.~\ref{sec:gatemon}, but this reduction is not significant and still permits high-fidelity gates for two qubits coupled by the nanowire.
We still choose the charging energies of each gatemon to be $E_{c1} = 255\ {\rm MHz}$ and $E_{c2} = 240\ {\rm MHz}$, as shown in Eq.~\eqref{eq:Parameters}. Below we choose parameters for beam splitters so that $K_{20, 02}\simeq 0.1$. For the specific case $K_{20, 02}= 0.123$, the qubit transition frequencies and anharmonicities are listed in Table~\ref{tbl:energies}. 
%\sout{Given the qubit frequencies and anharmonicities, we have}
%\begin{equation}
%\sout{	\dfrac{E_{11}}{h} = 12.53\ {\rm GHz}, \dfrac{E_{02}}{h} = 12.65\ {\rm GHz}, \dfrac{E_{20}}{h} = 12.27\ {\rm GHz}\label{eq:E110220_JQ}.}
%\end{equation}
For non-ideal transmission of conduction channels, the qubit frequency is smaller, but additional conduction channels in the junction can help adjust this frequency to a desirable value. 

Due to relatively large connectivity of the individual qubit junction, the absolute value of anharmonicities are suppressed so that the state $\widetilde{\ket{11}}$ is sandwiched between states $\widetilde{\ket{02}}$ and $\widetilde{\ket{20}}$. As expected from previously discussed two qubit systems, the non-zero transmission gives rise to finite $\Delta_{\rm CZ}$, shown by a thick solid line in Fig.~\ref{fig: energy_diagram_JQ}, enabling the construction of CZ gate by switching $T_c$. 
 
%\subsection{Gate fidelity}

\begin{figure}
	\centering
	\includegraphics[width=8cm]{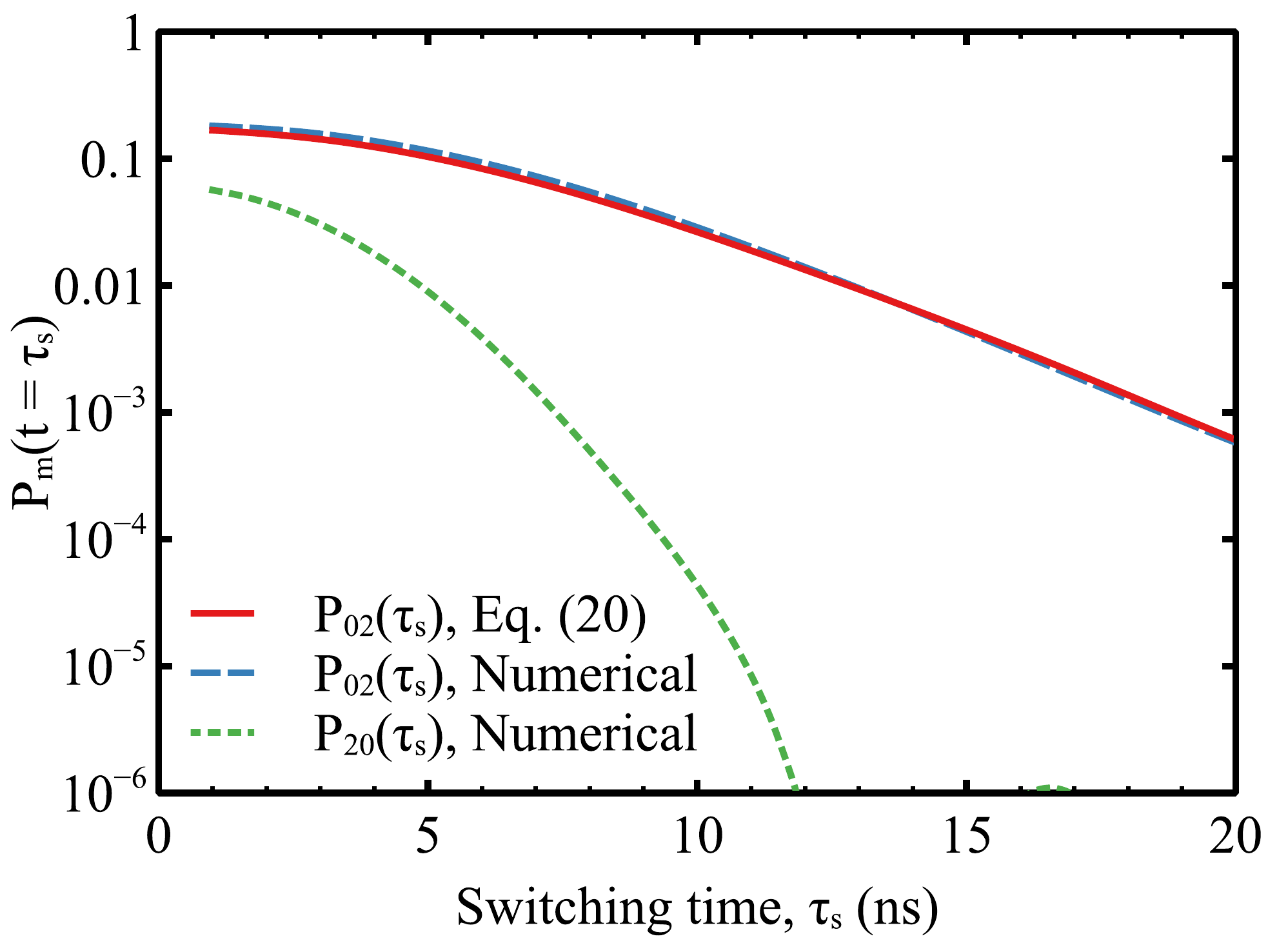}
	\caption{Transition probability to from state $\ket{11}$ to state $\ket{02}$ and $\ket{20}$ as a function of switching time for the H-junction qubits. "On"-transmission $T_0 = 0.6$. Charging energies and superconducting gap are the same as the transmon qubits and gatemon qubits. For our choice of parameters, $K_{02, 20} = 0.123$, corresponding to Josephson energy $E_J/h = 20.24\ {\rm GHz}$.}
	\label{fig:P02_Tswitch_JQ}
\end{figure}

In absence of decoherence, the gate error comes from the transitions to levels outside the computational basis during switching on/off interactions. Taking $T_0 = 0.6$ for the connecting wire, we investigate how transition probability decreases as  switching time $\tau_s$ increases. As shown in Fig.~\ref{fig:P02_Tswitch_JQ}, the transition error can be readily reduced to $10^{-3}$ for switching time longer than $20\ {\rm ns}$.
%========================================================================
%========================================================================

\begin{figure}
	\centering
	\includegraphics[width=8cm]{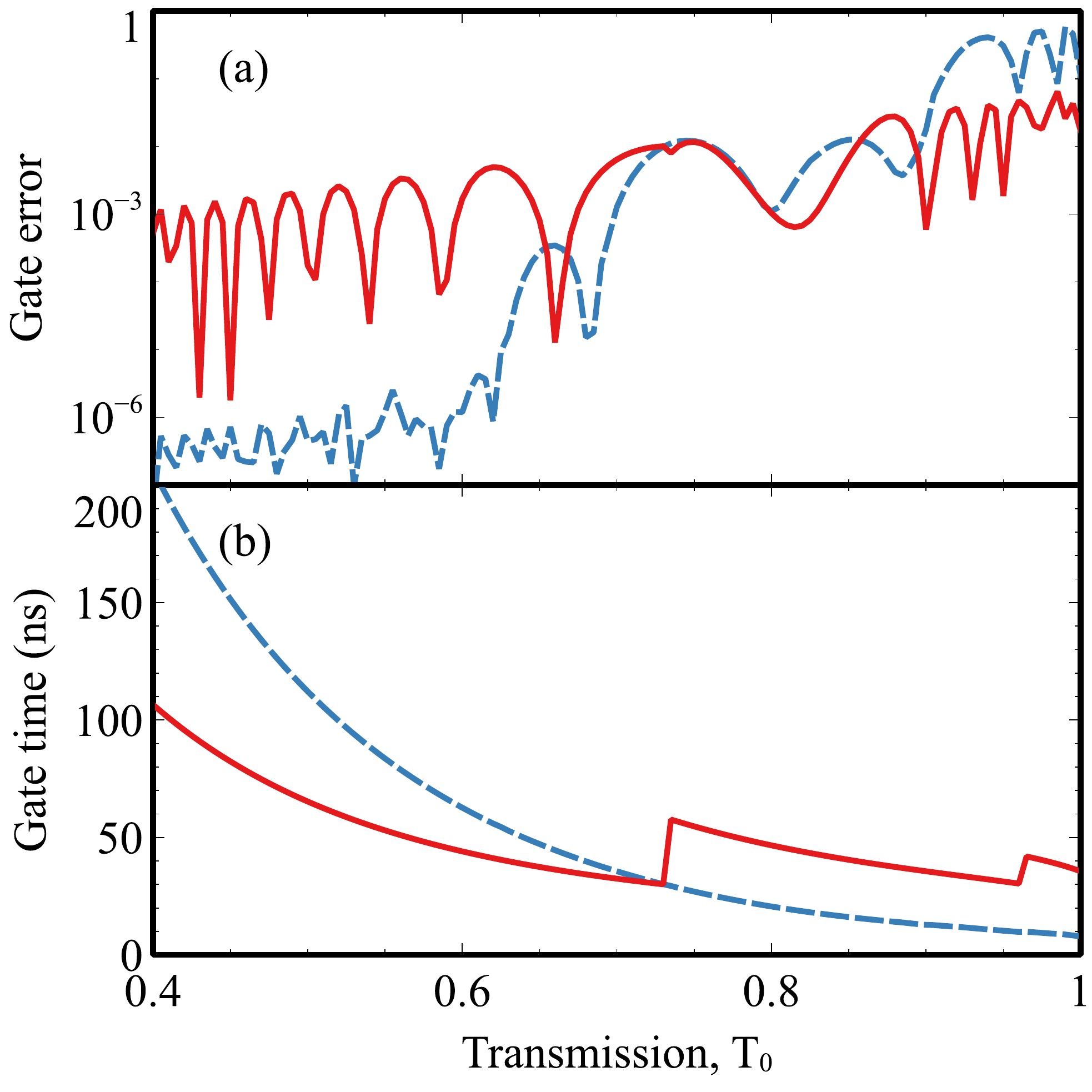}
	\caption{(a) (Color online) Error for the CZ gate and (b) corresponding gate time as functions of "on"-transmission $T_0$ of an H-pair. The solid red lines are calculated assuming a pulse shape with an interacting plateau, so the gate time is $\tau_w+2\tau_s$. Here, we take $\tau_s = 15\ {\rm ns}$. The dashed blue lines assumes the transmission is turned off right after it arrives it's maximum value $T_0$, thus the gate time is $2\tau_s$.}
	\label{fig:fid_transp_jq}
\end{figure}

The relatively small change of $\Delta_{\rm CZ}$ in Fig.~\ref{fig: energy_diagram_JQ} in transmission from $T_c=0$ to $T_c=0.8$ indicates a quite flexible tunable coupler for the H-pair. For a fixed switching time, smaller 'on'-interaction, which corresponds to smaller $T_0$, can prevent transitions during switching, but requires longer time to accomplish CZ gate. In contrast, larger interaction is good for gate time, though it acquires larger transition error. Truncating unpractical ranges for 'on'-transmission $T_0$, we make plots of gate error and time as functions of $T_0$, as shown in Fig.~\ref{fig:fid_transp_jq}. 
%========================================================================
%========================================================================
Figure~\ref{fig:fid_transp_jq} shows that for a large range of wire transmission, the worst case error can be smaller than or on the order of $10^{-2}$ with the gate time shorter than $200\ {\rm ns}$ for this parameter set.

\section{Conclusions\label{sec:conclusion}} 
\begin{figure}
	\centering
	\includegraphics[width=8cm]{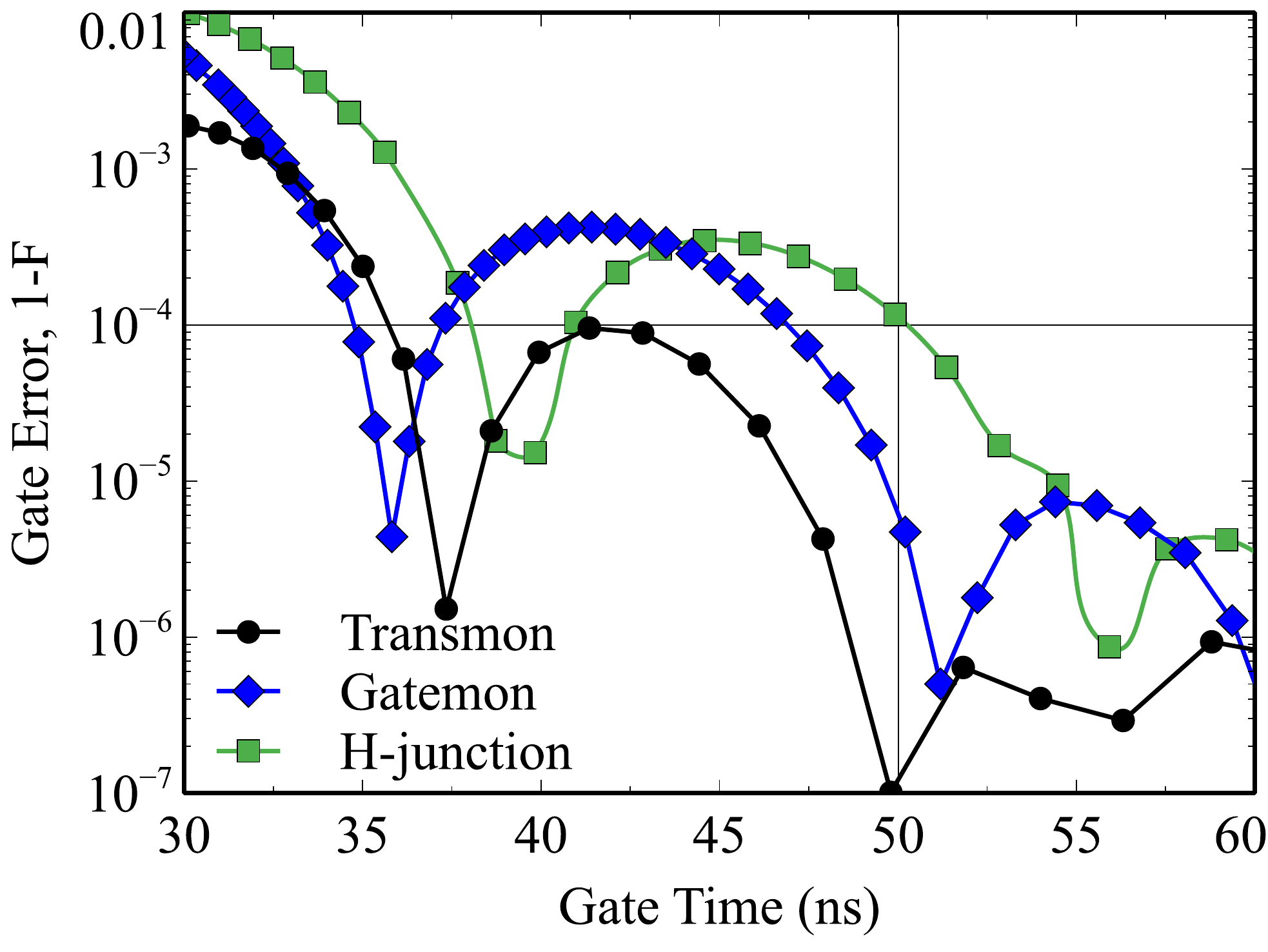}
	\caption{Error as a function of gate time for qubit systems analyzed above with waiting time $\tau_w = 0$. The charging energies are the same as what we used in previous figures. The Josephson energy $E_j = 20.55\ \rm GHz$ for transmons and gatemons, and $E_j = 20.24 \ \rm GHz$ for qubits in H-pair. The corresponding qubit frequencies are listed in Table.~\ref{tbl:energies}. For gate time about $50\ {\rm ns}$, all three qubit configurations allow for gate errors at the order of or less than $10^{-4}$.}
	\label{fig:comparison}
\end{figure}

We compared three kinds of qubits, namely a pair of transmons or gatemons as well as an H-pair by theoretically realizing a CZ gates through tuning inter-qubit interaction and taking a look at their gate fidelity and gate time as functions of experimentally practical transmission range. For all three systems, fidelity and gate time are generally not positively correlated and optimal transmission needs to be chosen to ensure reasonable gate time and error. Fortunately, numerical analysis show the existence of this optimal regime of transmission for all three systems, as shown in Fig.~\ref{fig:comparison}.

%========================================================================
%========================================================================

%For our choice of parameters, transmon possesses a larger $\Delta_{\rm CZ}$ so the gate time can be shorter even for the same qubit interaction. 
In general, a transmon has larger anharmonicities than a gatemon or H-pair.  This explains the different relative positions of states $\widetilde{\ket{11}}$ and $\widetilde{\ket{02}}$ for transmon system and the other two in this paper. For other choices of charging energies, this difference may not manifest as the obvious different energy configurations, but depending on different ways of realizing qubit gates, the transmon with larger anharmonicity may have a better fidelity as it distinguishes transitions among its computational basis from other transitions better than small anharmonicity system.

In Fig.~\ref{fig:comparison}, all three qubit systems demonstrate small error ($\le 10^{-4}$) for gate time about $50\ {\rm ns}$. The H-pair has longer gate time in our choice of parameter set. However, when the coupling between beam splitter and connecting junction is small, it allows a wide range of connector transmission for tuning the system. And when equipped with gatemon, it inherits the freedom of tuning qubit frequencies. Therefore, even with slightly longer gate time and potential operation error compared to traditional transmon, it allows for easy control of the energy scales and interactions, thus both single qubit gates and two-qubit entangling gates can be easily realized with little error coming from cross-talk or frequency crowding. 

The CZ gate can be realized in systems with always-on interaction by shifting qubits' spectrum in resonance to temporarily enhance the interaction between two qubits, \textit{e.g.} by bringing energies of states $\ket{11}$ and $\ket{20}$ to the same value~\cite{Strauch2003,Ghosh2013}. But even for the "off" configuration of qubits, the phase difference $\propto \Delta_{\rm CZ}$ continuously accumulates and reduces gate fidelities of single qubits. Such error accumulation becomes especially crucial for large scale qubit systems. Also, changing qubit frequency in the crowded spectrum of a large interacting system will cause numerous level crossing that will cumulatively result in large gate errors. Therefore, a tunable coupler between the qubits is a necessary element for a scalable quantum processor. The previous realization of inductive tunable coupler was utilized in g-mon systems\cite{Chen2014a,Geller2015,Neill2017}, where the coupling is controlled by the flux bias. In this paper we demonstrated that an inductive tunable coupling can also be controlled by electrostatic gate voltages by utilizing the epitaxial semiconductor Josephson junctions.

\begin{acknowledgements}
We are thankful to J. Chow, J. Ghosh, R. McDermott, K. Nesterov and I. Pechenezhskiy for fruitful discussions.  
The work of H.X. and A.L. was in part supported by NSF Grant No. DMR-1653661, UW-Madison Vilas Life Cycle Professorship program, and the Wisconsin Alumni Research Foundation.
\end{acknowledgements}
	 
\appendix*
	 
	 \section{Scattering matrix for the H-junction}
	 
	 We construct the scattering matrix $\hat{S}$ for the H-junction as shown in Fig.~\ref{fig:Hjunction}.
	 The short-wire scattering matrix $\hat{W}$ takes the form 
	 \be  \label{w-s}
	 \hat{W} = \begin{pmatrix} 
	 	r      &     t  \\
	 	t      &    r' 
	 \end{pmatrix}, 
	 \ee
	 with $r= \sqrt{1-T} e^{i \vartheta}$, $r'= \sqrt{1-T} e^{i (2\eta - \vartheta)}$ 
	 and $t = \sqrt{T} e^{i \eta}$, where $T$ is the transmission and $\vartheta$ and $\eta$ 
	 are two independent phases.

	 A beam-splitter scattering matrix $\hat{Y}^{(l,r)}$ is characterized by six parameters as follows: 
	 \begin{align} \label{s-tbs-g}
	 & Y_{11} = a \, e^{i \varphi_{11}}, \quad Y_{12} = b \sqrt{1-a^2} \, e^{i \varphi_{12}}, \nonumber \\
	 & Y_{13} = \sqrt{(1-a^2)(1-b^2)} \, e^{i \varphi_{13}}, \nonumber \\
	 & Y_{22} = -a b^2 e^{i(2\varphi_{12}-\varphi_{11})} + (1-b^2) e^{i \varphi_{22}}, \nonumber \\
	 & Y_{23} = -b\sqrt{1-b^2} \, e^{i \varphi_{13}} \left[ a  e^{i(\varphi_{12}-\varphi_{11})}+ e^{i(\varphi_{22}-\varphi_{12})} \right],\nonumber \\
	 & Y_{33} = e^{i 2 \varphi_{13}} \left[ -a (1-b^2) e^{-i \varphi_{11}} + b^2 e^{i(\varphi_{22} - 2 \varphi_{12})} \right], \nonumber \\
	 & Y_{\a\b} = Y_{\b \a}, \quad 1 \le \a < \b \le 3,
	 \end{align}
	 where $a,b \in [0,1]$, and $\varphi_{11,22,12,13} \in [0,2\pi]$. For $a=1$, $b=0$, and $b=1$, the 1, 2, and 3 lead is decoupled to the rest, respectively. The $H$-junction scattering matrix elements $S_{\a\b}$ are determined by the linear equations
%\be
%\begin{pmatrix} X_\a & S_{\a \b } & S_{\a+2, \b}  \end{pmatrix} = \begin{pmatrix} r X_{\a} + t X_{\bar{\a}} & \mathbf{e}_{\a,\b}  \end{pmatrix} \hat{Y}_\a,
%\ee
%\begin{subequations}
\begin{equation}
\begin{split}
\begin{pmatrix}
S_{00} \\ 
S_{10} \\ 
x
\end{pmatrix} & = \hat Y^{(l)} \begin{pmatrix}
1 \\ 
0\\ 
r x + t y 
\end{pmatrix},\,\,
\begin{pmatrix}
S_{20} \\ 
S_{30} \\ 
y
\end{pmatrix} = \hat Y^{(r)} \begin{pmatrix}
0 \\ 
0\\ 
t x + r' y 
\end{pmatrix},\\
%\end{equation}
%\begin{equation}
\begin{pmatrix}
S_{01} \\ 
S_{11} \\ 
x
\end{pmatrix} & = \hat Y^{(l)} \begin{pmatrix}
0 \\ 
1\\ 
r x + t y 
\end{pmatrix},\,\,
\begin{pmatrix}
S_{21} \\ 
S_{31} \\ 
y
\end{pmatrix} = \hat Y^{(r)} \begin{pmatrix}
0 \\ 
0\\ 
t x + r' y 
\end{pmatrix},\\
%\end{equation}
%\begin{equation}
\begin{pmatrix}
S_{22} \\ 
S_{32} \\ 
y
\end{pmatrix} & = \hat Y^{(r)} \begin{pmatrix}
1 \\ 
0\\ 
t x + r' y 
\end{pmatrix},\,\,
\begin{pmatrix}
S_{02} \\ 
S_{12} \\ 
x
\end{pmatrix} = \hat Y^{(l)} \begin{pmatrix}
0 \\ 
0\\ 
t y + r x
\end{pmatrix},\\
%\end{equation}
%\begin{equation}
\begin{pmatrix}
S_{23} \\ 
S_{33} \\ 
y
\end{pmatrix} & = \hat Y^{(r)} \begin{pmatrix}
0 \\ 
0\\ 
t x + r' y 
\end{pmatrix},\,\,
\begin{pmatrix}
S_{03} \\ 
S_{13} \\ 
x
\end{pmatrix} = \hat Y^{(l)} \begin{pmatrix}
0 \\ 
0\\ 
t y + r x 
\end{pmatrix}.
\end{split}
\end{equation}
%\end{subequations}

%where $\a \in \{ 1, 2\}$ indicting the two beam splitters, $\bar{1} (\bar{2}) \equiv 2 (1)$, and $\b \in \{1, 2, 3, 4 \}$. $X_\a$ are two unknown variables, $\hat{Y}_\a$ is determined by Eq.~(\ref{s-tbs-g}) with all parameters depending on $\a$, and the two-component vectors $\mathbf{e}_{1,1}=\mathbf{e}_{2,2}=(1, \, 0)$, $\mathbf{e}_{13}=\mathbf{e}_{24}=(0, \, 1)$, and $\mathbf{e}_{\a\b}=(0, \, 0)$, otherwise. 

%\bibliography{CZsemi}

\begin{thebibliography}{37}%
\makeatletter
\providecommand \@ifxundefined [1]{%
 \@ifx{#1\undefined}
}%
\providecommand \@ifnum [1]{%
 \ifnum #1\expandafter \@firstoftwo
 \else \expandafter \@secondoftwo
 \fi
}%
\providecommand \@ifx [1]{%
 \ifx #1\expandafter \@firstoftwo
 \else \expandafter \@secondoftwo
 \fi
}%
\providecommand \natexlab [1]{#1}%
\providecommand \enquote  [1]{``#1''}%
\providecommand \bibnamefont  [1]{#1}%
\providecommand \bibfnamefont [1]{#1}%
\providecommand \citenamefont [1]{#1}%
\providecommand \href@noop [0]{\@secondoftwo}%
\providecommand \href [0]{\begingroup \@sanitize@url \@href}%
\providecommand \@href[1]{\@@startlink{#1}\@@href}%
\providecommand \@@href[1]{\endgroup#1\@@endlink}%
\providecommand \@sanitize@url [0]{\catcode `\\12\catcode `\$12\catcode
  `\&12\catcode `\#12\catcode `\^12\catcode `\_12\catcode `\%12\relax}%
\providecommand \@@startlink[1]{}%
\providecommand \@@endlink[0]{}%
\providecommand \url  [0]{\begingroup\@sanitize@url \@url }%
\providecommand \@url [1]{\endgroup\@href {#1}{\urlprefix }}%
\providecommand \urlprefix  [0]{URL }%
\providecommand \Eprint [0]{\href }%
\providecommand \doibase [0]{http://dx.doi.org/}%
\providecommand \selectlanguage [0]{\@gobble}%
\providecommand \bibinfo  [0]{\@secondoftwo}%
\providecommand \bibfield  [0]{\@secondoftwo}%
\providecommand \translation [1]{[#1]}%
\providecommand \BibitemOpen [0]{}%
\providecommand \bibitemStop [0]{}%
\providecommand \bibitemNoStop [0]{.\EOS\space}%
\providecommand \EOS [0]{\spacefactor3000\relax}%
\providecommand \BibitemShut  [1]{\csname bibitem#1\endcsname}%
\let\auto@bib@innerbib\@empty
%</preamble>
\bibitem [{\citenamefont {Clarke}\ and\ \citenamefont
  {Wilhelm}(2008)}]{clarke2008superconducting}%
  \BibitemOpen
  \bibfield  {author} {\bibinfo {author} {\bibfnamefont {J.}~\bibnamefont
  {Clarke}}\ and\ \bibinfo {author} {\bibfnamefont {F.~K.}\ \bibnamefont
  {Wilhelm}},\ }\href@noop {} {\bibfield  {journal} {\bibinfo  {journal}
  {Nature}\ }\textbf {\bibinfo {volume} {453}},\ \bibinfo {pages} {1031}
  (\bibinfo {year} {2008})}\BibitemShut {NoStop}%
\bibitem [{\citenamefont {Lucero}\ \emph {et~al.}(2008)\citenamefont {Lucero},
  \citenamefont {Hofheinz}, \citenamefont {Ansmann}, \citenamefont {Bialczak},
  \citenamefont {Katz}, \citenamefont {Neeley}, \citenamefont {O'Connell},
  \citenamefont {Wang}, \citenamefont {Cleland},\ and\ \citenamefont
  {Martinis}}]{Lucero2008}%
  \BibitemOpen
  \bibfield  {author} {\bibinfo {author} {\bibfnamefont {E.}~\bibnamefont
  {Lucero}}, \bibinfo {author} {\bibfnamefont {M.}~\bibnamefont {Hofheinz}},
  \bibinfo {author} {\bibfnamefont {M.}~\bibnamefont {Ansmann}}, \bibinfo
  {author} {\bibfnamefont {R.~C.}\ \bibnamefont {Bialczak}}, \bibinfo {author}
  {\bibfnamefont {N.}~\bibnamefont {Katz}}, \bibinfo {author} {\bibfnamefont
  {M.}~\bibnamefont {Neeley}}, \bibinfo {author} {\bibfnamefont {A.~D.}\
  \bibnamefont {O'Connell}}, \bibinfo {author} {\bibfnamefont {H.}~\bibnamefont
  {Wang}}, \bibinfo {author} {\bibfnamefont {A.~N.}\ \bibnamefont {Cleland}}, \
  and\ \bibinfo {author} {\bibfnamefont {J.~M.}\ \bibnamefont {Martinis}},\
  }\href {\doibase 10.1103/PhysRevLett.100.247001} {\bibfield  {journal}
  {\bibinfo  {journal} {Phys. Rev. Lett.}\ }\textbf {\bibinfo {volume} {100}},\
  \bibinfo {pages} {247001} (\bibinfo {year} {2008})}\BibitemShut {NoStop}%
\bibitem [{\citenamefont {DiCarlo}\ \emph {et~al.}(2009)\citenamefont
  {DiCarlo}, \citenamefont {Chow}, \citenamefont {Gambetta}, \citenamefont
  {Bishop}, \citenamefont {Johnson}, \citenamefont {Schuster}, \citenamefont
  {Majer}, \citenamefont {Blais}, \citenamefont {Frunzio}, \citenamefont
  {Girvin} \emph {et~al.}}]{dicarlo2009demonstration}%
  \BibitemOpen
  \bibfield  {author} {\bibinfo {author} {\bibfnamefont {L.}~\bibnamefont
  {DiCarlo}}, \bibinfo {author} {\bibfnamefont {J.}~\bibnamefont {Chow}},
  \bibinfo {author} {\bibfnamefont {J.}~\bibnamefont {Gambetta}}, \bibinfo
  {author} {\bibfnamefont {L.~S.}\ \bibnamefont {Bishop}}, \bibinfo {author}
  {\bibfnamefont {B.}~\bibnamefont {Johnson}}, \bibinfo {author} {\bibfnamefont
  {D.}~\bibnamefont {Schuster}}, \bibinfo {author} {\bibfnamefont
  {J.}~\bibnamefont {Majer}}, \bibinfo {author} {\bibfnamefont
  {A.}~\bibnamefont {Blais}}, \bibinfo {author} {\bibfnamefont
  {L.}~\bibnamefont {Frunzio}}, \bibinfo {author} {\bibfnamefont
  {S.}~\bibnamefont {Girvin}},  \emph {et~al.},\ }\href@noop {} {\bibfield
  {journal} {\bibinfo  {journal} {Nature}\ }\textbf {\bibinfo {volume} {460}},\
  \bibinfo {pages} {240} (\bibinfo {year} {2009})}\BibitemShut {NoStop}%
\bibitem [{\citenamefont {Paik}\ \emph {et~al.}(2011)\citenamefont {Paik},
  \citenamefont {Schuster}, \citenamefont {Bishop}, \citenamefont {Kirchmair},
  \citenamefont {Catelani}, \citenamefont {Sears}, \citenamefont {Johnson},
  \citenamefont {Reagor}, \citenamefont {Frunzio}, \citenamefont {Glazman},
  \citenamefont {Girvin}, \citenamefont {Devoret},\ and\ \citenamefont
  {Schoelkopf}}]{Paik2011}%
  \BibitemOpen
  \bibfield  {author} {\bibinfo {author} {\bibfnamefont {H.}~\bibnamefont
  {Paik}}, \bibinfo {author} {\bibfnamefont {D.~I.}\ \bibnamefont {Schuster}},
  \bibinfo {author} {\bibfnamefont {L.~S.}\ \bibnamefont {Bishop}}, \bibinfo
  {author} {\bibfnamefont {G.}~\bibnamefont {Kirchmair}}, \bibinfo {author}
  {\bibfnamefont {G.}~\bibnamefont {Catelani}}, \bibinfo {author}
  {\bibfnamefont {A.~P.}\ \bibnamefont {Sears}}, \bibinfo {author}
  {\bibfnamefont {B.~R.}\ \bibnamefont {Johnson}}, \bibinfo {author}
  {\bibfnamefont {M.~J.}\ \bibnamefont {Reagor}}, \bibinfo {author}
  {\bibfnamefont {L.}~\bibnamefont {Frunzio}}, \bibinfo {author} {\bibfnamefont
  {L.~I.}\ \bibnamefont {Glazman}}, \bibinfo {author} {\bibfnamefont {S.~M.}\
  \bibnamefont {Girvin}}, \bibinfo {author} {\bibfnamefont {M.~H.}\
  \bibnamefont {Devoret}}, \ and\ \bibinfo {author} {\bibfnamefont {R.~J.}\
  \bibnamefont {Schoelkopf}},\ }\href {\doibase 10.1103/PhysRevLett.107.240501}
  {\bibfield  {journal} {\bibinfo  {journal} {Phys. Rev. Lett.}\ }\textbf
  {\bibinfo {volume} {107}},\ \bibinfo {pages} {240501} (\bibinfo {year}
  {2011})}\BibitemShut {NoStop}%
\bibitem [{\citenamefont {Devoret}\ and\ \citenamefont
  {Schoelkopf}(2013)}]{devoret2013superconducting}%
  \BibitemOpen
  \bibfield  {author} {\bibinfo {author} {\bibfnamefont {M.~H.}\ \bibnamefont
  {Devoret}}\ and\ \bibinfo {author} {\bibfnamefont {R.~J.}\ \bibnamefont
  {Schoelkopf}},\ }\href@noop {} {\bibfield  {journal} {\bibinfo  {journal}
  {Science}\ }\textbf {\bibinfo {volume} {339}},\ \bibinfo {pages} {1169}
  (\bibinfo {year} {2013})}\BibitemShut {NoStop}%
\bibitem [{\citenamefont {Wendin}(2017)}]{wendin2017quantum}%
  \BibitemOpen
  \bibfield  {author} {\bibinfo {author} {\bibfnamefont {G.}~\bibnamefont
  {Wendin}},\ }\href@noop {} {\bibfield  {journal} {\bibinfo  {journal}
  {Reports on Progress in Physics}\ }\textbf {\bibinfo {volume} {80}},\
  \bibinfo {pages} {106001} (\bibinfo {year} {2017})}\BibitemShut {NoStop}%
\bibitem [{\citenamefont {Yamamoto}\ \emph {et~al.}(2003)\citenamefont
  {Yamamoto}, \citenamefont {Pashkin}, \citenamefont {Astafiev}, \citenamefont
  {Nakamura},\ and\ \citenamefont {Tsai}}]{yamamoto2003demonstration}%
  \BibitemOpen
  \bibfield  {author} {\bibinfo {author} {\bibfnamefont {T.}~\bibnamefont
  {Yamamoto}}, \bibinfo {author} {\bibfnamefont {Y.~A.}\ \bibnamefont
  {Pashkin}}, \bibinfo {author} {\bibfnamefont {O.}~\bibnamefont {Astafiev}},
  \bibinfo {author} {\bibfnamefont {Y.}~\bibnamefont {Nakamura}}, \ and\
  \bibinfo {author} {\bibfnamefont {J.-S.}\ \bibnamefont {Tsai}},\ }\href@noop
  {} {\bibfield  {journal} {\bibinfo  {journal} {Nature}\ }\textbf {\bibinfo
  {volume} {425}},\ \bibinfo {pages} {941} (\bibinfo {year}
  {2003})}\BibitemShut {NoStop}%
\bibitem [{\citenamefont {Chow}\ \emph {et~al.}(2011)\citenamefont {Chow},
  \citenamefont {C\'orcoles}, \citenamefont {Gambetta}, \citenamefont
  {Rigetti}, \citenamefont {Johnson}, \citenamefont {Smolin}, \citenamefont
  {Rozen}, \citenamefont {Keefe}, \citenamefont {Rothwell}, \citenamefont
  {Ketchen},\ and\ \citenamefont {Steffen}}]{chow2011simple}%
  \BibitemOpen
  \bibfield  {author} {\bibinfo {author} {\bibfnamefont {J.~M.}\ \bibnamefont
  {Chow}}, \bibinfo {author} {\bibfnamefont {A.~D.}\ \bibnamefont
  {C\'orcoles}}, \bibinfo {author} {\bibfnamefont {J.~M.}\ \bibnamefont
  {Gambetta}}, \bibinfo {author} {\bibfnamefont {C.}~\bibnamefont {Rigetti}},
  \bibinfo {author} {\bibfnamefont {B.~R.}\ \bibnamefont {Johnson}}, \bibinfo
  {author} {\bibfnamefont {J.~A.}\ \bibnamefont {Smolin}}, \bibinfo {author}
  {\bibfnamefont {J.~R.}\ \bibnamefont {Rozen}}, \bibinfo {author}
  {\bibfnamefont {G.~A.}\ \bibnamefont {Keefe}}, \bibinfo {author}
  {\bibfnamefont {M.~B.}\ \bibnamefont {Rothwell}}, \bibinfo {author}
  {\bibfnamefont {M.~B.}\ \bibnamefont {Ketchen}}, \ and\ \bibinfo {author}
  {\bibfnamefont {M.}~\bibnamefont {Steffen}},\ }\href {\doibase
  10.1103/PhysRevLett.107.080502} {\bibfield  {journal} {\bibinfo  {journal}
  {Phys. Rev. Lett.}\ }\textbf {\bibinfo {volume} {107}},\ \bibinfo {pages}
  {080502} (\bibinfo {year} {2011})}\BibitemShut {NoStop}%
\bibitem [{\citenamefont {Plantenberg}\ \emph {et~al.}(2007)\citenamefont
  {Plantenberg}, \citenamefont {de~Groot}, \citenamefont {Harmans},\ and\
  \citenamefont {Mooij}}]{Plantenberg2007}%
  \BibitemOpen
  \bibfield  {author} {\bibinfo {author} {\bibfnamefont {J.~H.}\ \bibnamefont
  {Plantenberg}}, \bibinfo {author} {\bibfnamefont {P.~C.}\ \bibnamefont
  {de~Groot}}, \bibinfo {author} {\bibfnamefont {C.~J. P.~M.}\ \bibnamefont
  {Harmans}}, \ and\ \bibinfo {author} {\bibfnamefont {J.~E.}\ \bibnamefont
  {Mooij}},\ }\href {http://dx.doi.org/10.1038/nature05896} {\bibfield
  {journal} {\bibinfo  {journal} {Nature}\ }\textbf {\bibinfo {volume} {447}},\
  \bibinfo {pages} {836 EP } (\bibinfo {year} {2007})}\BibitemShut {NoStop}%
\bibitem [{\citenamefont {Leek}\ \emph {et~al.}(2009)\citenamefont {Leek},
  \citenamefont {Filipp}, \citenamefont {Maurer}, \citenamefont {Baur},
  \citenamefont {Bianchetti}, \citenamefont {Fink}, \citenamefont {G\"oppl},
  \citenamefont {Steffen},\ and\ \citenamefont {Wallraff}}]{Leek2009}%
  \BibitemOpen
  \bibfield  {author} {\bibinfo {author} {\bibfnamefont {P.~J.}\ \bibnamefont
  {Leek}}, \bibinfo {author} {\bibfnamefont {S.}~\bibnamefont {Filipp}},
  \bibinfo {author} {\bibfnamefont {P.}~\bibnamefont {Maurer}}, \bibinfo
  {author} {\bibfnamefont {M.}~\bibnamefont {Baur}}, \bibinfo {author}
  {\bibfnamefont {R.}~\bibnamefont {Bianchetti}}, \bibinfo {author}
  {\bibfnamefont {J.~M.}\ \bibnamefont {Fink}}, \bibinfo {author}
  {\bibfnamefont {M.}~\bibnamefont {G\"oppl}}, \bibinfo {author} {\bibfnamefont
  {L.}~\bibnamefont {Steffen}}, \ and\ \bibinfo {author} {\bibfnamefont
  {A.}~\bibnamefont {Wallraff}},\ }\href {\doibase 10.1103/PhysRevB.79.180511}
  {\bibfield  {journal} {\bibinfo  {journal} {Phys. Rev. B}\ }\textbf {\bibinfo
  {volume} {79}},\ \bibinfo {pages} {180511} (\bibinfo {year}
  {2009})}\BibitemShut {NoStop}%
\bibitem [{\citenamefont {Rigetti}\ and\ \citenamefont
  {Devoret}(2010)}]{Rigetti2010}%
  \BibitemOpen
  \bibfield  {author} {\bibinfo {author} {\bibfnamefont {C.}~\bibnamefont
  {Rigetti}}\ and\ \bibinfo {author} {\bibfnamefont {M.}~\bibnamefont
  {Devoret}},\ }\href {\doibase 10.1103/PhysRevB.81.134507} {\bibfield
  {journal} {\bibinfo  {journal} {Phys. Rev. B}\ }\textbf {\bibinfo {volume}
  {81}},\ \bibinfo {pages} {134507} (\bibinfo {year} {2010})}\BibitemShut
  {NoStop}%
\bibitem [{\citenamefont {Sheldon}\ \emph {et~al.}(2016)\citenamefont
  {Sheldon}, \citenamefont {Magesan}, \citenamefont {Chow},\ and\ \citenamefont
  {Gambetta}}]{Sheldon2016}%
  \BibitemOpen
  \bibfield  {author} {\bibinfo {author} {\bibfnamefont {S.}~\bibnamefont
  {Sheldon}}, \bibinfo {author} {\bibfnamefont {E.}~\bibnamefont {Magesan}},
  \bibinfo {author} {\bibfnamefont {J.~M.}\ \bibnamefont {Chow}}, \ and\
  \bibinfo {author} {\bibfnamefont {J.~M.}\ \bibnamefont {Gambetta}},\ }\href
  {\doibase 10.1103/PhysRevA.93.060302} {\bibfield  {journal} {\bibinfo
  {journal} {Phys. Rev. A}\ }\textbf {\bibinfo {volume} {93}},\ \bibinfo
  {pages} {060302} (\bibinfo {year} {2016})}\BibitemShut {NoStop}%
\bibitem [{\citenamefont {Chang}\ \emph {et~al.}(2013)\citenamefont {Chang},
  \citenamefont {Manucharyan}, \citenamefont {Jespersen}, \citenamefont
  {Nyg{\aa}rd},\ and\ \citenamefont {Marcus}}]{chang2013tunneling}%
  \BibitemOpen
  \bibfield  {author} {\bibinfo {author} {\bibfnamefont {W.}~\bibnamefont
  {Chang}}, \bibinfo {author} {\bibfnamefont {V.E.}~\bibnamefont {Manucharyan}},
  \bibinfo {author} {\bibfnamefont {T.S.}~\bibnamefont {Jespersen}}, \bibinfo
  {author} {\bibfnamefont {J.}~\bibnamefont {Nyg{\aa}rd}}, \ and\ \bibinfo
  {author} {\bibfnamefont {C.~M.}\ \bibnamefont {Marcus}},\ }\href@noop {}
  {\bibfield  {journal} {\bibinfo  {journal} {Phys. Rev. Lett.}\ }\textbf
  {\bibinfo {volume} {110}},\ \bibinfo {pages} {217005} (\bibinfo {year}
  {2013})}\BibitemShut {NoStop}%
\bibitem [{\citenamefont {Larsen}\ \emph {et~al.}(2015)\citenamefont {Larsen},
  \citenamefont {Petersson}, \citenamefont {Kuemmeth}, \citenamefont
  {Jespersen}, \citenamefont {Krogstrup}, \citenamefont {Nyg\aa{}rd},\ and\
  \citenamefont {Marcus}}]{Lars2015}%
  \BibitemOpen
  \bibfield  {author} {\bibinfo {author} {\bibfnamefont {T.~W.}\ \bibnamefont
  {Larsen}}, \bibinfo {author} {\bibfnamefont {K.~D.}\ \bibnamefont
  {Petersson}}, \bibinfo {author} {\bibfnamefont {F.}~\bibnamefont {Kuemmeth}},
  \bibinfo {author} {\bibfnamefont {T.~S.}\ \bibnamefont {Jespersen}}, \bibinfo
  {author} {\bibfnamefont {P.}~\bibnamefont {Krogstrup}}, \bibinfo {author}
  {\bibfnamefont {J.}~\bibnamefont {Nyg\aa{}rd}}, \ and\ \bibinfo {author}
  {\bibfnamefont {C.~M.}\ \bibnamefont {Marcus}},\ }\href {\doibase
  10.1103/PhysRevLett.115.127001} {\bibfield  {journal} {\bibinfo  {journal}
  {Phys. Rev. Lett.}\ }\textbf {\bibinfo {volume} {115}},\ \bibinfo {pages}
  {127001} (\bibinfo {year} {2015})}\BibitemShut {NoStop}%
\bibitem [{\citenamefont {de~Lange}\ \emph {et~al.}(2015)\citenamefont
  {de~Lange}, \citenamefont {van Heck}, \citenamefont {Bruno}, \citenamefont
  {van Woerkom}, \citenamefont {Geresdi}, \citenamefont {Plissard},
  \citenamefont {Bakkers}, \citenamefont {Akhmerov},\ and\ \citenamefont
  {DiCarlo}}]{DeLange2015}%
  \BibitemOpen
  \bibfield  {author} {\bibinfo {author} {\bibfnamefont {G.}~\bibnamefont
  {de~Lange}}, \bibinfo {author} {\bibfnamefont {B.}~\bibnamefont {van Heck}},
  \bibinfo {author} {\bibfnamefont {A.}~\bibnamefont {Bruno}}, \bibinfo
  {author} {\bibfnamefont {D.~J.}\ \bibnamefont {van Woerkom}}, \bibinfo
  {author} {\bibfnamefont {A.}~\bibnamefont {Geresdi}}, \bibinfo {author}
  {\bibfnamefont {S.~R.}\ \bibnamefont {Plissard}}, \bibinfo {author}
  {\bibfnamefont {E.~P. A.~M.}\ \bibnamefont {Bakkers}}, \bibinfo {author}
  {\bibfnamefont {A.~R.}\ \bibnamefont {Akhmerov}}, \ and\ \bibinfo {author}
  {\bibfnamefont {L.}~\bibnamefont {DiCarlo}},\ }\href {\doibase
  10.1103/PhysRevLett.115.127002} {\bibfield  {journal} {\bibinfo  {journal}
  {Phys. Rev. Lett.}\ }\textbf {\bibinfo {volume} {115}},\ \bibinfo {pages}
  {127002} (\bibinfo {year} {2015})}\BibitemShut {NoStop}%
\bibitem [{\citenamefont {Shabani}\ \emph {et~al.}(2016)\citenamefont
  {Shabani}, \citenamefont {Kjaergaard}, \citenamefont {Suominen},
  \citenamefont {Kim}, \citenamefont {Nichele}, \citenamefont {Pakrouski},
  \citenamefont {Stankevic}, \citenamefont {Lutchyn}, \citenamefont
  {Krogstrup}, \citenamefont {Feidenhans'l}, \citenamefont {Kraemer},
  \citenamefont {Nayak}, \citenamefont {Troyer}, \citenamefont {Marcus},\ and\
  \citenamefont {Palmstr\o{}m}}]{Shabani2016}%
  \BibitemOpen
  \bibfield  {author} {\bibinfo {author} {\bibfnamefont {J.}~\bibnamefont
  {Shabani}}, \bibinfo {author} {\bibfnamefont {M.}~\bibnamefont {Kjaergaard}},
  \bibinfo {author} {\bibfnamefont {H.~J.}\ \bibnamefont {Suominen}}, \bibinfo
  {author} {\bibfnamefont {Y.}~\bibnamefont {Kim}}, \bibinfo {author}
  {\bibfnamefont {F.}~\bibnamefont {Nichele}}, \bibinfo {author} {\bibfnamefont
  {K.}~\bibnamefont {Pakrouski}}, \bibinfo {author} {\bibfnamefont
  {T.}~\bibnamefont {Stankevic}}, \bibinfo {author} {\bibfnamefont {R.~M.}\
  \bibnamefont {Lutchyn}}, \bibinfo {author} {\bibfnamefont {P.}~\bibnamefont
  {Krogstrup}}, \bibinfo {author} {\bibfnamefont {R.}~\bibnamefont
  {Feidenhans'l}}, \bibinfo {author} {\bibfnamefont {S.}~\bibnamefont
  {Kraemer}}, \bibinfo {author} {\bibfnamefont {C.}~\bibnamefont {Nayak}},
  \bibinfo {author} {\bibfnamefont {M.}~\bibnamefont {Troyer}}, \bibinfo
  {author} {\bibfnamefont {C.~M.}\ \bibnamefont {Marcus}}, \ and\ \bibinfo
  {author} {\bibfnamefont {C.~J.}\ \bibnamefont {Palmstr\o{}m}},\ }\href
  {\doibase 10.1103/PhysRevB.93.155402} {\bibfield  {journal} {\bibinfo
  {journal} {Phys. Rev. B}\ }\textbf {\bibinfo {volume} {93}},\ \bibinfo
  {pages} {155402} (\bibinfo {year} {2016})}\BibitemShut {NoStop}%
\bibitem [{\citenamefont {Doh}\ \emph {et~al.}(2005)\citenamefont {Doh},
  \citenamefont {van Dam}, \citenamefont {Roest}, \citenamefont {Bakkers},
  \citenamefont {Kouwenhoven},\ and\ \citenamefont {De~Franceschi}}]{Doh272}%
  \BibitemOpen
  \bibfield  {author} {\bibinfo {author} {\bibfnamefont {Y.-J.}\ \bibnamefont
  {Doh}}, \bibinfo {author} {\bibfnamefont {J.~A.}\ \bibnamefont {van Dam}},
  \bibinfo {author} {\bibfnamefont {A.~L.}\ \bibnamefont {Roest}}, \bibinfo
  {author} {\bibfnamefont {E.~P. A.~M.}\ \bibnamefont {Bakkers}}, \bibinfo
  {author} {\bibfnamefont {L.~P.}\ \bibnamefont {Kouwenhoven}}, \ and\ \bibinfo
  {author} {\bibfnamefont {S.}~\bibnamefont {De~Franceschi}},\ }\href {\doibase
  10.1126/science.1113523} {\bibfield  {journal} {\bibinfo  {journal}
  {Science}\ }\textbf {\bibinfo {volume} {309}},\ \bibinfo {pages} {272}
  (\bibinfo {year} {2005})}\BibitemShut {NoStop}%
\bibitem [{\citenamefont {Casparis}\ \emph {et~al.}(2016)\citenamefont
  {Casparis}, \citenamefont {Larsen}, \citenamefont {Olsen}, \citenamefont
  {Kuemmeth}, \citenamefont {Krogstrup}, \citenamefont {Nyg{\aa}rd},
  \citenamefont {Petersson},\ and\ \citenamefont {Marcus}}]{Casparis2016}%
  \BibitemOpen
  \bibfield  {author} {\bibinfo {author} {\bibfnamefont {L.}~\bibnamefont
  {Casparis}}, \bibinfo {author} {\bibfnamefont {T.~W.}\ \bibnamefont
  {Larsen}}, \bibinfo {author} {\bibfnamefont {M.~S.}\ \bibnamefont {Olsen}},
  \bibinfo {author} {\bibfnamefont {F.}~\bibnamefont {Kuemmeth}}, \bibinfo
  {author} {\bibfnamefont {P.}~\bibnamefont {Krogstrup}}, \bibinfo {author}
  {\bibfnamefont {J.}~\bibnamefont {Nyg{\aa}rd}}, \bibinfo {author}
  {\bibfnamefont {K.~D.}\ \bibnamefont {Petersson}}, \ and\ \bibinfo {author}
  {\bibfnamefont {C.~M.}\ \bibnamefont {Marcus}},\ }\href {\doibase
  10.1103/PhysRevLett.116.150505} {\bibfield  {journal} {\bibinfo  {journal}
  {Phys. Rev. Lett.}\ }\textbf {\bibinfo {volume} {116}},\ \bibinfo {pages}
  {150505} (\bibinfo {year} {2016})}\BibitemShut {NoStop}%
\bibitem [{\citenamefont {Kringh{\o}j}\ \emph {et~al.}(2017)\citenamefont
  {Kringh{\o}j}, \citenamefont {Casparis}, \citenamefont {Hell}, \citenamefont
  {Larsen}, \citenamefont {Kuemmeth}, \citenamefont {Leijnse}, \citenamefont
  {Flensberg}, \citenamefont {Krogstrup}, \citenamefont {Nyg{\aa}rd},
  \citenamefont {Petersson},\ and\ \citenamefont {Marcus}}]{Kringhoj2017}%
  \BibitemOpen
  \bibfield  {author} {\bibinfo {author} {\bibfnamefont {A.}~\bibnamefont
  {Kringh{\o}j}}, \bibinfo {author} {\bibfnamefont {L.}~\bibnamefont
  {Casparis}}, \bibinfo {author} {\bibfnamefont {M.}~\bibnamefont {Hell}},
  \bibinfo {author} {\bibfnamefont {T.~W.}\ \bibnamefont {Larsen}}, \bibinfo
  {author} {\bibfnamefont {F.}~\bibnamefont {Kuemmeth}}, \bibinfo {author}
  {\bibfnamefont {M.}~\bibnamefont {Leijnse}}, \bibinfo {author} {\bibfnamefont
  {K.}~\bibnamefont {Flensberg}}, \bibinfo {author} {\bibfnamefont
  {P.}~\bibnamefont {Krogstrup}}, \bibinfo {author} {\bibfnamefont
  {J.}~\bibnamefont {Nyg{\aa}rd}}, \bibinfo {author} {\bibfnamefont {K.~D.}\
  \bibnamefont {Petersson}}, \ and\ \bibinfo {author} {\bibfnamefont {C.~M.}\
  \bibnamefont {Marcus}},\ }\href {http://arxiv.org/abs/1703.05643} {\bibfield
  {journal} {\bibinfo  {journal} {arXiv:1703.05643}\ } (\bibinfo {year}
  {2017})}\BibitemShut {NoStop}%
\bibitem [{\citenamefont {Strauch}\ \emph {et~al.}(2003)\citenamefont
  {Strauch}, \citenamefont {Johnson}, \citenamefont {Dragt}, \citenamefont
  {Lobb}, \citenamefont {Anderson},\ and\ \citenamefont
  {Wellstood}}]{Strauch2003}%
  \BibitemOpen
  \bibfield  {author} {\bibinfo {author} {\bibfnamefont {F.~W.}\ \bibnamefont
  {Strauch}}, \bibinfo {author} {\bibfnamefont {P.~R.}\ \bibnamefont
  {Johnson}}, \bibinfo {author} {\bibfnamefont {A.~J.}\ \bibnamefont {Dragt}},
  \bibinfo {author} {\bibfnamefont {C.~J.}\ \bibnamefont {Lobb}}, \bibinfo
  {author} {\bibfnamefont {J.~R.}\ \bibnamefont {Anderson}}, \ and\ \bibinfo
  {author} {\bibfnamefont {F.~C.}\ \bibnamefont {Wellstood}},\ }\href {\doibase
  10.1103/PhysRevLett.91.167005} {\bibfield  {journal} {\bibinfo  {journal}
  {Phys. Rev. Lett.}\ }\textbf {\bibinfo {volume} {91}},\ \bibinfo {pages}
  {167005} (\bibinfo {year} {2003})}\BibitemShut {NoStop}%
\bibitem [{\citenamefont {Bialczak}\ \emph {et~al.}(2010)\citenamefont
  {Bialczak}, \citenamefont {Ansmann}, \citenamefont {Hofheinz}, \citenamefont
  {Lucero}, \citenamefont {Neeley}, \citenamefont {O'Connell}, \citenamefont
  {Sank}, \citenamefont {Wang}, \citenamefont {Wenner}, \citenamefont
  {Steffen}, \citenamefont {Cleland},\ and\ \citenamefont
  {Martinis}}]{Bialczak2010}%
  \BibitemOpen
  \bibfield  {author} {\bibinfo {author} {\bibfnamefont {R.~C.}\ \bibnamefont
  {Bialczak}}, \bibinfo {author} {\bibfnamefont {M.}~\bibnamefont {Ansmann}},
  \bibinfo {author} {\bibfnamefont {M.}~\bibnamefont {Hofheinz}}, \bibinfo
  {author} {\bibfnamefont {E.}~\bibnamefont {Lucero}}, \bibinfo {author}
  {\bibfnamefont {M.}~\bibnamefont {Neeley}}, \bibinfo {author} {\bibfnamefont
  {A.~D.}\ \bibnamefont {O'Connell}}, \bibinfo {author} {\bibfnamefont
  {D.}~\bibnamefont {Sank}}, \bibinfo {author} {\bibfnamefont {H.}~\bibnamefont
  {Wang}}, \bibinfo {author} {\bibfnamefont {J.}~\bibnamefont {Wenner}},
  \bibinfo {author} {\bibfnamefont {M.}~\bibnamefont {Steffen}}, \bibinfo
  {author} {\bibfnamefont {A.~N.}\ \bibnamefont {Cleland}}, \ and\ \bibinfo
  {author} {\bibfnamefont {J.~M.}\ \bibnamefont {Martinis}},\ }\href {\doibase
  10.1038/nphys1639} {\bibfield  {journal} {\bibinfo  {journal} {Nat. Phys.}\
  }\textbf {\bibinfo {volume} {6}},\ \bibinfo {pages} {409} (\bibinfo {year}
  {2010})}\BibitemShut {NoStop}%
\bibitem [{\citenamefont {Yamamoto}\ \emph {et~al.}(2010)\citenamefont
  {Yamamoto}, \citenamefont {Neeley}, \citenamefont {Lucero}, \citenamefont
  {Bialczak}, \citenamefont {Kelly}, \citenamefont {Lenander}, \citenamefont
  {Mariantoni}, \citenamefont {O'Connell}, \citenamefont {Sank}, \citenamefont
  {Wang}, \citenamefont {Weides}, \citenamefont {Wenner}, \citenamefont {Yin},
  \citenamefont {Cleland},\ and\ \citenamefont {Martinis}}]{Yamamoto2010}%
  \BibitemOpen
  \bibfield  {author} {\bibinfo {author} {\bibfnamefont {T.}~\bibnamefont
  {Yamamoto}}, \bibinfo {author} {\bibfnamefont {M.}~\bibnamefont {Neeley}},
  \bibinfo {author} {\bibfnamefont {E.}~\bibnamefont {Lucero}}, \bibinfo
  {author} {\bibfnamefont {R.~C.}\ \bibnamefont {Bialczak}}, \bibinfo {author}
  {\bibfnamefont {J.}~\bibnamefont {Kelly}}, \bibinfo {author} {\bibfnamefont
  {M.}~\bibnamefont {Lenander}}, \bibinfo {author} {\bibfnamefont
  {M.}~\bibnamefont {Mariantoni}}, \bibinfo {author} {\bibfnamefont {A.~D.}\
  \bibnamefont {O'Connell}}, \bibinfo {author} {\bibfnamefont {D.}~\bibnamefont
  {Sank}}, \bibinfo {author} {\bibfnamefont {H.}~\bibnamefont {Wang}}, \bibinfo
  {author} {\bibfnamefont {M.}~\bibnamefont {Weides}}, \bibinfo {author}
  {\bibfnamefont {J.}~\bibnamefont {Wenner}}, \bibinfo {author} {\bibfnamefont
  {Y.}~\bibnamefont {Yin}}, \bibinfo {author} {\bibfnamefont {A.~N.}\
  \bibnamefont {Cleland}}, \ and\ \bibinfo {author} {\bibfnamefont {J.~M.}\
  \bibnamefont {Martinis}},\ }\href {\doibase 10.1103/PhysRevB.82.184515}
  {\bibfield  {journal} {\bibinfo  {journal} {Phys. Rev. B}\ }\textbf {\bibinfo
  {volume} {82}},\ \bibinfo {pages} {184515} (\bibinfo {year} {2010})}\BibitemShut
  {NoStop}%
\bibitem [{\citenamefont {Chen}\ \emph {et~al.}(2014)\citenamefont {Chen},
  \citenamefont {Neill}, \citenamefont {Roushan}, \citenamefont {Leung},
  \citenamefont {Fang}, \citenamefont {Barends}, \citenamefont {Kelly},
  \citenamefont {Campbell}, \citenamefont {Chen}, \citenamefont {Chiaro},
  \citenamefont {Dunsworth}, \citenamefont {Jeffrey}, \citenamefont {Megrant},
  \citenamefont {Mutus}, \citenamefont {O'Malley}, \citenamefont {Quintana},
  \citenamefont {Sank}, \citenamefont {Vainsencher}, \citenamefont {Wenner},
  \citenamefont {White}, \citenamefont {Geller}, \citenamefont {Cleland},\ and\
  \citenamefont {Martinis}}]{Chen2014a}%
  \BibitemOpen
  \bibfield  {author} {\bibinfo {author} {\bibfnamefont {Y.}~\bibnamefont
  {Chen}}, \bibinfo {author} {\bibfnamefont {C.}~\bibnamefont {Neill}},
  \bibinfo {author} {\bibfnamefont {P.}~\bibnamefont {Roushan}}, \bibinfo
  {author} {\bibfnamefont {N.}~\bibnamefont {Leung}}, \bibinfo {author}
  {\bibfnamefont {M.}~\bibnamefont {Fang}}, \bibinfo {author} {\bibfnamefont
  {R.}~\bibnamefont {Barends}}, \bibinfo {author} {\bibfnamefont
  {J.}~\bibnamefont {Kelly}}, \bibinfo {author} {\bibfnamefont
  {B.}~\bibnamefont {Campbell}}, \bibinfo {author} {\bibfnamefont
  {Z.}~\bibnamefont {Chen}}, \bibinfo {author} {\bibfnamefont {B.}~\bibnamefont
  {Chiaro}}, \bibinfo {author} {\bibfnamefont {A.}~\bibnamefont {Dunsworth}},
  \bibinfo {author} {\bibfnamefont {E.}~\bibnamefont {Jeffrey}}, \bibinfo
  {author} {\bibfnamefont {A.}~\bibnamefont {Megrant}}, \bibinfo {author}
  {\bibfnamefont {J.~Y.}\ \bibnamefont {Mutus}}, \bibinfo {author}
  {\bibfnamefont {P.~J.J.}\ \bibnamefont {O'Malley}}, \bibinfo {author}
  {\bibfnamefont {C.~M.}\ \bibnamefont {Quintana}}, \bibinfo {author}
  {\bibfnamefont {D.}~\bibnamefont {Sank}}, \bibinfo {author} {\bibfnamefont
  {A.}~\bibnamefont {Vainsencher}}, \bibinfo {author} {\bibfnamefont
  {J.}~\bibnamefont {Wenner}}, \bibinfo {author} {\bibfnamefont {T.~C.}\
  \bibnamefont {White}}, \bibinfo {author} {\bibfnamefont {M.~R.}\ \bibnamefont
  {Geller}}, \bibinfo {author} {\bibfnamefont {A.~N.}\ \bibnamefont {Cleland}},
  \ and\ \bibinfo {author} {\bibfnamefont {J.~M.}\ \bibnamefont {Martinis}},\
  }\href {\doibase 10.1103/PhysRevLett.113.220502} {\bibfield  {journal}
  {\bibinfo  {journal} {Phys. Rev. Lett.}\ }\textbf {\bibinfo {volume} {113}},\
  \bibinfo {pages} {220502} (\bibinfo {year} {2014})}\BibitemShut {NoStop}%
\bibitem [{\citenamefont {Geller}\ \emph {et~al.}(2015)\citenamefont {Geller},
  \citenamefont {Donate}, \citenamefont {Chen}, \citenamefont {Fang},
  \citenamefont {Leung}, \citenamefont {Neill}, \citenamefont {Roushan},\ and\
  \citenamefont {Martinis}}]{Geller2015}%
  \BibitemOpen
  \bibfield  {author} {\bibinfo {author} {\bibfnamefont {M.~R.}\ \bibnamefont
  {Geller}}, \bibinfo {author} {\bibfnamefont {E.}~\bibnamefont {Donate}},
  \bibinfo {author} {\bibfnamefont {Y.}~\bibnamefont {Chen}}, \bibinfo {author}
  {\bibfnamefont {M.~T.}\ \bibnamefont {Fang}}, \bibinfo {author}
  {\bibfnamefont {N.}~\bibnamefont {Leung}}, \bibinfo {author} {\bibfnamefont
  {C.}~\bibnamefont {Neill}}, \bibinfo {author} {\bibfnamefont
  {P.}~\bibnamefont {Roushan}}, \ and\ \bibinfo {author} {\bibfnamefont
  {J.~M.}\ \bibnamefont {Martinis}},\ }\href {\doibase
  10.1103/PhysRevA.92.012320} {\bibfield  {journal} {\bibinfo  {journal} {Phys.
  Rev. A}\ }\textbf {\bibinfo {volume} {92}},\ \bibinfo {pages} {012320}
  (\bibinfo {year} {2015})}\BibitemShut {NoStop}%
\bibitem [{\citenamefont {Koch}\ \emph {et~al.}(2007)\citenamefont {Koch},
  \citenamefont {Yu}, \citenamefont {Gambetta}, \citenamefont {Houck},
  \citenamefont {Schuster}, \citenamefont {Majer}, \citenamefont {Blais},
  \citenamefont {Devoret}, \citenamefont {Girvin},\ and\ \citenamefont
  {Schoelkopf}}]{Koch2007}%
  \BibitemOpen
  \bibfield  {author} {\bibinfo {author} {\bibfnamefont {J.}~\bibnamefont
  {Koch}}, \bibinfo {author} {\bibfnamefont {T.M.}~\bibnamefont {Yu}}, \bibinfo
  {author} {\bibfnamefont {J.}~\bibnamefont {Gambetta}}, \bibinfo {author}
  {\bibfnamefont {A.A.}~\bibnamefont {Houck}}, \bibinfo {author} {\bibfnamefont
  {D.I.}~\bibnamefont {Schuster}}, \bibinfo {author} {\bibfnamefont
  {J.}~\bibnamefont {Majer}}, \bibinfo {author} {\bibfnamefont
  {A.}~\bibnamefont {Blais}}, \bibinfo {author} {\bibfnamefont
  {M.H.}~\bibnamefont {Devoret}}, \bibinfo {author} {\bibfnamefont
  {S.M.}~\bibnamefont {Girvin}}, \ and\ \bibinfo {author} {\bibfnamefont
  {R.J.}~\bibnamefont {Schoelkopf}},\ }\href {\doibase
  10.1103/PhysRevA.76.042319} {\bibfield  {journal} {\bibinfo  {journal} {Phys.
  Rev. A}\ }\textbf {\bibinfo {volume} {76}},\ \bibinfo {pages} {042319}
  (\bibinfo {year} {2007})}\BibitemShut {NoStop}%
\bibitem [{\citenamefont {Majer}\ \emph {et~al.}(2007)\citenamefont {Majer},
  \citenamefont {Chow}, \citenamefont {Gambetta}, \citenamefont {Koch},
  \citenamefont {Johnson}, \citenamefont {Schreier}, \citenamefont {Frunzio},
  \citenamefont {Schuster}, \citenamefont {Houck}, \citenamefont {Wallraff},
  \citenamefont {Blais}, \citenamefont {Devoret}, \citenamefont {Girvin},\ and\
  \citenamefont {Schoelkopf}}]{majer2007coupling}%
  \BibitemOpen
  \bibfield  {author} {\bibinfo {author} {\bibfnamefont {J.}~\bibnamefont
  {Majer}}, \bibinfo {author} {\bibfnamefont {J.}~\bibnamefont {Chow}},
  \bibinfo {author} {\bibfnamefont {J.}~\bibnamefont {Gambetta}}, \bibinfo
  {author} {\bibfnamefont {J.}~\bibnamefont {Koch}}, \bibinfo {author}
  {\bibfnamefont {B.}~\bibnamefont {Johnson}}, \bibinfo {author} {\bibfnamefont
  {J.}~\bibnamefont {Schreier}}, \bibinfo {author} {\bibfnamefont
  {L.}~\bibnamefont {Frunzio}}, \bibinfo {author} {\bibfnamefont
  {D.}~\bibnamefont {Schuster}}, \bibinfo {author} {\bibfnamefont
  {A.}~\bibnamefont {Houck}}, \bibinfo {author} {\bibfnamefont
  {A.}~\bibnamefont {Wallraff}}, \bibinfo {author} {\bibfnamefont
  {A.}~\bibnamefont {Blais}}, \bibinfo {author} {\bibfnamefont
  {M.}~\bibnamefont {Devoret}}, \bibinfo {author} {\bibfnamefont
  {S.}~\bibnamefont {Girvin}}, \ and\ \bibinfo {author} {\bibfnamefont
  {R.}~\bibnamefont {Schoelkopf}},\ }\href {\doibase 10.1038/nature06184}
  {\bibfield  {journal} {\bibinfo  {journal} {Nature}\ }\textbf {\bibinfo
  {volume} {449}},\ \bibinfo {pages} {443} (\bibinfo {year}
  {2007})}\BibitemShut {NoStop}%
\bibitem [{\citenamefont {Rigetti}\ \emph {et~al.}(2012)\citenamefont
  {Rigetti}, \citenamefont {Gambetta}, \citenamefont {Poletto}, \citenamefont
  {Plourde}, \citenamefont {Chow}, \citenamefont {C\'orcoles}, \citenamefont
  {Smolin}, \citenamefont {Merkel}, \citenamefont {Rozen}, \citenamefont
  {Keefe}, \citenamefont {Rothwell}, \citenamefont {Ketchen},\ and\
  \citenamefont {Steffen}}]{rigetti2012superconducting}%
  \BibitemOpen
  \bibfield  {author} {\bibinfo {author} {\bibfnamefont {C.}~\bibnamefont
  {Rigetti}}, \bibinfo {author} {\bibfnamefont {J.~M.}\ \bibnamefont
  {Gambetta}}, \bibinfo {author} {\bibfnamefont {S.}~\bibnamefont {Poletto}},
  \bibinfo {author} {\bibfnamefont {B.~L.~T.}\ \bibnamefont {Plourde}},
  \bibinfo {author} {\bibfnamefont {J.~M.}\ \bibnamefont {Chow}}, \bibinfo
  {author} {\bibfnamefont {A.~D.}\ \bibnamefont {C\'orcoles}}, \bibinfo
  {author} {\bibfnamefont {J.~A.}\ \bibnamefont {Smolin}}, \bibinfo {author}
  {\bibfnamefont {S.~T.}\ \bibnamefont {Merkel}}, \bibinfo {author}
  {\bibfnamefont {J.~R.}\ \bibnamefont {Rozen}}, \bibinfo {author}
  {\bibfnamefont {G.~A.}\ \bibnamefont {Keefe}}, \bibinfo {author}
  {\bibfnamefont {M.~B.}\ \bibnamefont {Rothwell}}, \bibinfo {author}
  {\bibfnamefont {M.~B.}\ \bibnamefont {Ketchen}}, \ and\ \bibinfo {author}
  {\bibfnamefont {M.}~\bibnamefont {Steffen}},\ }\href {\doibase
  10.1103/PhysRevB.86.100506} {\bibfield  {journal} {\bibinfo  {journal} {Phys.
  Rev. B}\ }\textbf {\bibinfo {volume} {86}},\ \bibinfo {pages} {100506}
  (\bibinfo {year} {2012})}\BibitemShut {NoStop}%
\bibitem [{\citenamefont {van Heck}\ \emph {et~al.}(2014)\citenamefont {van
  Heck}, \citenamefont {Mi},\ and\ \citenamefont {Akhmerov}}]{van2014single}%
  \BibitemOpen
  \bibfield  {author} {\bibinfo {author} {\bibfnamefont {B.}~\bibnamefont {van
  Heck}}, \bibinfo {author} {\bibfnamefont {S.}~\bibnamefont {Mi}}, \ and\
  \bibinfo {author} {\bibfnamefont {A.~R.}\ \bibnamefont {Akhmerov}},\ }\href
  {\doibase 10.1103/PhysRevB.90.155450} {\bibfield  {journal} {\bibinfo
  {journal} {Phys. Rev. B}\ }\textbf {\bibinfo {volume} {90}},\ \bibinfo
  {pages} {155450} (\bibinfo {year} {2014})}\BibitemShut {NoStop}%
\bibitem [{\citenamefont {Riwar}\ \emph {et~al.}(2016)\citenamefont {Riwar},
  \citenamefont {Houzet}, \citenamefont {Meyer},\ and\ \citenamefont
  {Nazarov}}]{Nazarov-NC16}%
  \BibitemOpen
  \bibfield  {author} {\bibinfo {author} {\bibfnamefont {R.-P.}\ \bibnamefont
  {Riwar}}, \bibinfo {author} {\bibfnamefont {M.}~\bibnamefont {Houzet}},
  \bibinfo {author} {\bibfnamefont {J.~S.}\ \bibnamefont {Meyer}}, \ and\
  \bibinfo {author} {\bibfnamefont {Y.~V.}\ \bibnamefont {Nazarov}},\ }\href
  {http://dx.doi.org/10.1038/ncomms11167} {\bibfield  {journal} {\bibinfo
  {journal} {Nature Communications}\ }\textbf {\bibinfo {volume} {7}},\
  \bibinfo {pages} {11167 EP } (\bibinfo {year} {2016})}\BibitemShut {NoStop}%
\bibitem [{\citenamefont {Meyer}\ and\ \citenamefont
  {Houzet}(2017)}]{Meyer2017}%
  \BibitemOpen
  \bibfield  {author} {\bibinfo {author} {\bibfnamefont {J.~S.}\ \bibnamefont
  {Meyer}}\ and\ \bibinfo {author} {\bibfnamefont {M.}~\bibnamefont {Houzet}},\
  }\href {\doibase 10.1103/PhysRevLett.119.136807} {\bibfield  {journal}
  {\bibinfo  {journal} {Phys. Rev. Lett.}\ }\textbf {\bibinfo {volume} {119}},\
  \bibinfo {pages} {136807} (\bibinfo {year} {2017})}\BibitemShut {NoStop}%
\bibitem [{\citenamefont {Xie}\ \emph {et~al.}(2017{\natexlab{a}})\citenamefont
  {Xie}, \citenamefont {Vavilov},\ and\ \citenamefont {Levchenko}}]{Xie2017}%
  \BibitemOpen
  \bibfield  {author} {\bibinfo {author} {\bibfnamefont {H.-Y.}\ \bibnamefont
  {Xie}}, \bibinfo {author} {\bibfnamefont {M.~G.}\ \bibnamefont {Vavilov}}, \
  and\ \bibinfo {author} {\bibfnamefont {A.}~\bibnamefont {Levchenko}},\
  }\href@noop {} {\bibfield  {journal} {\bibinfo  {journal} {arXiv:1707.07712}\
  } (\bibinfo {year} {2017}{\natexlab{a}})}\BibitemShut {NoStop}%
\bibitem [{\citenamefont {Xie}\ \emph {et~al.}(2017{\natexlab{b}})\citenamefont
  {Xie}, \citenamefont {Vavilov},\ and\ \citenamefont {Levchenko}}]{Xie2017a}%
  \BibitemOpen
  \bibfield  {author} {\bibinfo {author} {\bibfnamefont {H.-Y.}\ \bibnamefont
  {Xie}}, \bibinfo {author} {\bibfnamefont {M.~G.}\ \bibnamefont {Vavilov}}, \
  and\ \bibinfo {author} {\bibfnamefont {A.}~\bibnamefont {Levchenko}},\ }\href
  {\doibase 10.1103/PhysRevB.96.161406} {\bibfield  {journal} {\bibinfo
  {journal} {Phys. Rev. B}\ }\textbf {\bibinfo {volume} {96}},\ \bibinfo
  {pages} {161406} (\bibinfo {year} {2017}{\natexlab{b}})}\BibitemShut
  {NoStop}%
\bibitem [{\citenamefont {Beenakker}(1991)}]{Beenakker-PRL91}%
  \BibitemOpen
  \bibfield  {author} {\bibinfo {author} {\bibfnamefont {C.~W.~J.}\
  \bibnamefont {Beenakker}},\ }\href {\doibase 10.1103/PhysRevLett.67.3836}
  {\bibfield  {journal} {\bibinfo  {journal} {Phys. Rev. Lett.}\ }\textbf
  {\bibinfo {volume} {67}},\ \bibinfo {pages} {3836} (\bibinfo {year}
  {1991})}\BibitemShut {NoStop}%
\bibitem [{\citenamefont {Chow}\ \emph {et~al.}(2013)\citenamefont {Chow},
  \citenamefont {Gambetta}, \citenamefont {Cross}, \citenamefont {Merkel},
  \citenamefont {Rigetti},\ and\ \citenamefont {Steffen}}]{Chow2013}%
  \BibitemOpen
  \bibfield  {author} {\bibinfo {author} {\bibfnamefont {J.~M.}\ \bibnamefont
  {Chow}}, \bibinfo {author} {\bibfnamefont {J.~M.}\ \bibnamefont {Gambetta}},
  \bibinfo {author} {\bibfnamefont {A.~W.}\ \bibnamefont {Cross}}, \bibinfo
  {author} {\bibfnamefont {S.~T.}\ \bibnamefont {Merkel}}, \bibinfo {author}
  {\bibfnamefont {C.}~\bibnamefont {Rigetti}}, \ and\ \bibinfo {author}
  {\bibfnamefont {M.}~\bibnamefont {Steffen}},\ }\href {\doibase
  10.1088/1367-2630/15/11/115012} {\bibfield  {journal} {\bibinfo  {journal}
  {New J. Phys.}\ }\textbf {\bibinfo {volume} {15}},\ \bibinfo {pages} {115012}
  (\bibinfo {year} {2013})}\BibitemShut {NoStop}%
\bibitem [{\citenamefont {Ghosh}\ \emph {et~al.}(2013)\citenamefont {Ghosh},
  \citenamefont {Galiautdinov}, \citenamefont {Zhou}, \citenamefont {Korotkov},
  \citenamefont {Martinis},\ and\ \citenamefont {Geller}}]{Ghosh2013}%
  \BibitemOpen
  \bibfield  {author} {\bibinfo {author} {\bibfnamefont {J.}~\bibnamefont
  {Ghosh}}, \bibinfo {author} {\bibfnamefont {A.}~\bibnamefont {Galiautdinov}},
  \bibinfo {author} {\bibfnamefont {Z.}~\bibnamefont {Zhou}}, \bibinfo {author}
  {\bibfnamefont {A.~N.}\ \bibnamefont {Korotkov}}, \bibinfo {author}
  {\bibfnamefont {J.~M.}\ \bibnamefont {Martinis}}, \ and\ \bibinfo {author}
  {\bibfnamefont {M.~R.}\ \bibnamefont {Geller}},\ }\href {\doibase
  10.1103/PhysRevA.87.022309} {\bibfield  {journal} {\bibinfo  {journal} {Phys.
  Rev. A}\ }\textbf {\bibinfo {volume} {87}},\ \bibinfo {pages} {022309}
  (\bibinfo {year} {2013})}\BibitemShut {NoStop}%
\bibitem [{\citenamefont {Pedersen}\ \emph {et~al.}(2007)\citenamefont
  {Pedersen}, \citenamefont {M{\o}ller},\ and\ \citenamefont
  {M{\o}lmer}}]{Pedersen2007}%
  \BibitemOpen
  \bibfield  {author} {\bibinfo {author} {\bibfnamefont {L.~H.}\ \bibnamefont
  {Pedersen}}, \bibinfo {author} {\bibfnamefont {N.~M.}\ \bibnamefont
  {M{\o}ller}}, \ and\ \bibinfo {author} {\bibfnamefont {K.}~\bibnamefont
  {M{\o}lmer}},\ }\href@noop {} {\bibfield  {journal} {\bibinfo  {journal}
  {Physics Letters A}\ }\textbf {\bibinfo {volume} {367}},\ \bibinfo {pages}
  {47} (\bibinfo {year} {2007})}\BibitemShut {NoStop}%
\bibitem [{\citenamefont {Neill}\ \emph {et~al.}(2017)\citenamefont {Neill},
  \citenamefont {Roushan}, \citenamefont {Kechedzhi}, \citenamefont {Boixo},
  \citenamefont {Isakov}, \citenamefont {Smelyanskiy}, \citenamefont {Barends},
  \citenamefont {Burkett}, \citenamefont {Chen}, \citenamefont {Chen},
  \citenamefont {Chiaro}, \citenamefont {Dunsworth}, \citenamefont {Fowler},
  \citenamefont {Foxen}, \citenamefont {Graff}, \citenamefont {Jeffrey},
  \citenamefont {Kelly}, \citenamefont {Lucero}, \citenamefont {Megrant},
  \citenamefont {Mutus}, \citenamefont {Neeley}, \citenamefont {Quintana},
  \citenamefont {Sank}, \citenamefont {Vainsencher}, \citenamefont {Wenner},
  \citenamefont {White}, \citenamefont {Neven},\ and\ \citenamefont
  {Martinis}}]{Neill2017}%
  \BibitemOpen
  \bibfield  {author} {\bibinfo {author} {\bibfnamefont {C.}~\bibnamefont
  {Neill}}, \bibinfo {author} {\bibfnamefont {P.}~\bibnamefont {Roushan}},
  \bibinfo {author} {\bibfnamefont {K.}~\bibnamefont {Kechedzhi}}, \bibinfo
  {author} {\bibfnamefont {S.}~\bibnamefont {Boixo}}, \bibinfo {author}
  {\bibfnamefont {S.~V.}\ \bibnamefont {Isakov}}, \bibinfo {author}
  {\bibfnamefont {V.}~\bibnamefont {Smelyanskiy}}, \bibinfo {author}
  {\bibfnamefont {R.}~\bibnamefont {Barends}}, \bibinfo {author} {\bibfnamefont
  {B.}~\bibnamefont {Burkett}}, \bibinfo {author} {\bibfnamefont
  {Y.}~\bibnamefont {Chen}}, \bibinfo {author} {\bibfnamefont {Z.}~\bibnamefont
  {Chen}}, \bibinfo {author} {\bibfnamefont {B.}~\bibnamefont {Chiaro}},
  \bibinfo {author} {\bibfnamefont {A.}~\bibnamefont {Dunsworth}}, \bibinfo
  {author} {\bibfnamefont {A.}~\bibnamefont {Fowler}}, \bibinfo {author}
  {\bibfnamefont {B.}~\bibnamefont {Foxen}}, \bibinfo {author} {\bibfnamefont
  {R.}~\bibnamefont {Graff}}, \bibinfo {author} {\bibfnamefont
  {E.}~\bibnamefont {Jeffrey}}, \bibinfo {author} {\bibfnamefont
  {J.}~\bibnamefont {Kelly}}, \bibinfo {author} {\bibfnamefont
  {E.}~\bibnamefont {Lucero}}, \bibinfo {author} {\bibfnamefont
  {A.}~\bibnamefont {Megrant}}, \bibinfo {author} {\bibfnamefont
  {J.}~\bibnamefont {Mutus}}, \bibinfo {author} {\bibfnamefont
  {M.}~\bibnamefont {Neeley}}, \bibinfo {author} {\bibfnamefont
  {C.}~\bibnamefont {Quintana}}, \bibinfo {author} {\bibfnamefont
  {D.}~\bibnamefont {Sank}}, \bibinfo {author} {\bibfnamefont {A.}~\bibnamefont
  {Vainsencher}}, \bibinfo {author} {\bibfnamefont {J.}~\bibnamefont {Wenner}},
  \bibinfo {author} {\bibfnamefont {T.~C.}\ \bibnamefont {White}}, \bibinfo
  {author} {\bibfnamefont {H.}~\bibnamefont {Neven}}, \ and\ \bibinfo {author}
  {\bibfnamefont {J.~M.}\ \bibnamefont {Martinis}},\ }\href
  {http://arxiv.org/abs/1709.06678} {\  (\bibinfo {year} {2017})},\ \Eprint
  {http://arxiv.org/abs/1709.06678} {arXiv:1709.06678} \BibitemShut {NoStop}%
\end{thebibliography}

%merlin.mbs apsrev4-1.bst 2010-07-25 4.21a (PWD, AO, DPC) hacked
%Control: key (0)
%Control: author (72) initials jnrlst
%Control: editor formatted (1) identically to author
%Control: production of article title (-1) disabled
%Control: page (0) single
%Control: year (1) truncated
%Control: production of eprint (0) enabled
%

\end{document}